\pdfoutput=1

\documentclass[12pt]{iopart}

\usepackage{amscd,amssymb,a4}
\usepackage{graphicx,xspace}
\expandafter\let\csname equation*\endcsname\relax
\expandafter\let\csname endequation*\endcsname\relax
\usepackage{amsmath}
\usepackage{float}

\usepackage[left=2cm,top=2cm,right=2cm,bottom=2cm]{geometry}
\usepackage{color}

\begin{document}

\newcommand{\red}{\color{red}}
\newcommand{\blue}{\color{blue}}
\newcommand{\todo}[1]{\textbf{To do: #1}}
\newcommand{\beq}{\begin{equation}}
\newcommand{\eeq}{\end{equation}}
\newcommand{\bea}{\begin{eqnarray}}
\newcommand{\eea}{\end{eqnarray}}
\newcommand{\rar}{\rightarrow}
\newcommand{\lar}{\leftarrow}
\newcommand{\ra}{\right\rangle}
\newcommand{\la}{\left\langle }
\renewcommand{\d}{{\rm d }}
\newcommand{\m}{{\tilde m }}
\newcommand{\p}{\partial}
\newcommand{\nn}{\nonumber }

\newcommand{\fig}[2]{\includegraphics[width=#1]{./figures/#2}}
\newcommand{\Fig}[1]{\includegraphics[width=7cm]{./figures/#1}}
\newlength{\bilderlength}
\newcommand{\bilderscale}{0.35}
\newcommand{\storebilderscale}{\bilderscale}
\newcommand{\bilderskip}{\hspace*{0.8ex}}
\newcommand{\textdiagram}[1]{%
\renewcommand{\bilderscale}{0.2}%
\diagram{#1}\renewcommand{\bilderscale}{\storebilderscale}}
\newcommand{\vardiagram}[2]{%
\newcommand{\bilderscale}{#1}%
\diagram{#2}\renewcommand{\bilderscale}{\storebilderscale}}
\newcommand{\diagram}[1]{%
\settowidth{\bilderlength}{\bilderskip%
\includegraphics[scale=\bilderscale]{./figures/#1}\bilderskip}%
\parbox{\bilderlength}{\bilderskip%
\includegraphics[scale=\bilderscale]{./figures/#1}\bilderskip}}
\newcommand{\Diagram}[1]{%
\settowidth{\bilderlength}{%
\includegraphics[scale=\bilderscale]{./figures/#1}}%
\parbox{\bilderlength}{%
\includegraphics[scale=\bilderscale]{./figures/#1}}}
\bibliographystyle{KAY}

%


%








\newcommand{\atanh}
{\operatorname{atanh}}

\newcommand{\ArcTan}
{\operatorname{ArcTan}}

\newcommand{\ArcCoth}
{\operatorname{ArcCoth}}

\newcommand{\Erf}
{\operatorname{Erf}}

\newcommand{\Erfi}
{\operatorname{Erfi}}

\newcommand{\Ei}
{\operatorname{Ei}}

\newcommand{\sgn}{{\mathrm{sgn}}}
\def\be{\begin{equation}}
\def\ee{\end{equation}}

\def\bea{\begin{eqnarray}}
\def\eea{\end{eqnarray}}

\def\e{\epsilon}
\def\l{\lambda}
\def\d{\delta}
\def\o{\omega}
\def\cb{\bar{c}}
\def\Li{{\rm Li}}

\title[Log-Gamma polymer and replica]{Log-Gamma directed polymer with fixed endpoints via the replica Bethe Ansatz}

\author{Thimoth\'ee Thiery and Pierre Le Doussal}

\address{CNRS-Laboratoire de Physique Th\'eorique de l'Ecole Normale Sup\'erieure\\
24 rue Lhomond, 75231 Paris
Cedex-France}



%



\date{\today}

\begin{abstract}
 We study the model of a discrete directed polymer (DP) on the square lattice with homogeneous inverse gamma
distribution of site random Boltzmann weights, introduced by Seppalainen \cite{logsep1}. 
The integer moments of the partition sum, $\overline{Z^n}$, are studied using a transfer
matrix formulation, which appears as a generalization of the Lieb-Liniger quantum mechanics 
of bosons to discrete time and space. In the present case of the inverse gamma
distribution the model is integrable in terms of a coordinate Bethe Ansatz, as discovered
by Brunet. Using the Brunet-Bethe eigenstates we obtain an exact expression for the
integer moments of $\overline{Z^n}$ for polymers of arbitrary lengths and fixed endpoint positions. 
Although these moments do not exist for all integer $n$, we are nevertheless able to construct
a generating function which reproduces all existing integer moments, and which takes
the form of a Fredholm determinant (FD). This suggests an analytic continuation via a Mellin-Barnes 
transform and we thereby propose a FD ansatz representation for the probability
distribution function (PDF) of $Z$ and its Laplace transform. In the limit of very long DP, this ansatz yields that the
distribution of the free energy converges to the Gaussian unitary ensemble (GUE) Tracy-Widom distribution
up to a non-trivial average and variance that we calculate. Our asymptotic predictions 
coincide with a result by Borodin et al. \cite{logboro} based on a formula obtained by Corwin et al. \cite{logsep2} using the geometric Robinson-Schensted-Knuth (gRSK) correspondence. In addition we obtain the dependence on the endpoint position and the exact elastic coefficient at large time. 
We argue the equivalence between our formula and the one of Borodin et al.
As we discuss, this provides connections between 
quantum integrability and tropical combinatorics.
\end{abstract}

\maketitle

\newpage

\section{Introduction}\label{Introduction}

Recently it was realized that methods of integrability in quantum systems could 
be used to obtain exact solutions for the one dimensional continuum Kardar-Parisi-Zhang equation (KPZ).
The KPZ equation \cite{KPZ} is 
a paradigmatic model for 1D noisy growth processes, encompassing a vast universality class of discrete growth 
or equivalent models (the so-called KPZ class). The probability distribution function (PDF) of the
KPZ height field $h$ at time $t$ was obtained (at one, or several space points) and shown to converge at large $t$ to the universal Tracy-Widom
(TW) distributions \cite{TW1994} for the largest eigenvalues of large Gaussian random matrices. 

One route, entirely within continuum models, is to use the Cole-Hopf mapping onto the
problem of the directed polymer, $h \sim \ln Z$, where $h$ is the height of the KPZ interface,
and $Z$ the partition sum (in the statistical mechanics sense) of continuum directed paths in presence of quenched
disorder. Using the replica method, the time evolution of the moments $\overline{Z^n}$ 
maps \cite{kardareplica} onto the (imaginary time) quantum evolution of bosons with attractive interactions, the
so-called Lieb-Liniger model \cite{ll}. This model is integrable via the Bethe Ansatz, which ultimately
yields exact expressions for the integer moments $\overline{Z^n}$ of $P(Z)$, the PDF of $Z$. 
Although recovering from there the PDF of the KPZ height field requires the use of some heuristics
(since the moments actually grow too fast to ensure uniqueness), this method
allowed to obtain the Laplace transform of $P(Z)$ (also called generating function)
for all the important classes of KPZ initial conditions (droplet, flat, stationary, half-space)
\cite{we,dotsenko,we-flat,SasamotoStationary,dg-12,d-13,psn-11,ld-14,cl-14}.
Interestingly, in all the solvable cases, it was obtained as a Fredholm determinant,
with various kernels and valid for all times $t$. Let us also mention the recently observed connection between the continuum model and the sine-Gordon quantum field theory \cite{sineG}.

Another route is to study appropriate discrete models, which, in some
limit, reproduce the continuum result. This route is favored in the mathematics community
since it does not suffer, in the favorable cases, from the moment problem. In 
\cite{spohnKPZEdge,corwinDP,reviewCorwin}, the solution for the continuum KPZ 
equation with droplet initial conditions was obtained 
as the weak asymmetry limit of the ASEP. Another integrable discrete model, the
$q$-TASEP, also exhibits such a limit for $q \to 1$, and was shown to be 
part of a broader integrability structure related to Macdonald processes.
This allows for  rigorous extensions to the other class of KPZ initial conditions, 
which are under intense current scrutiny \cite{BorodinMacdo,BorodinQboson,BCS,BCF,quastel-prep}. 

Among the solvable discrete models, are the discrete and semi-discrete 
directed polymer models. The model studied by Johansson in \cite{Johansson2000} considers a DP on 
a square lattice with a geometric distribution of 
the on-site random potentials, and allows for an exact solution. It is
a zero temperature DP model since it focuses on the path with minimal energy
(energy being additive along a path), as in the last passage percolation models. 
Another remarkable solvable model is called the log-gamma polymer
and was introduced by Seppalainen \cite{logsep1}. 
It is a finite temperature model as it focuses on Boltzmann weights (which are multiplicative
along a path). Its peculiarity is that the random weights on the sites are distributed 
according to a so-called inverse gamma distribution, which has a power law
fat tail. Such a choice for the quenched disorder leads to remarkable properties: 
an exact expression for the Laplace transform of $P(Z)$ (the generating function) 
was obtained by Corwin et al. in \cite{logsep2}. 
The method is quite involved and uses combinatorics methods known as the gRSK correspondence
(a geometric lifting of the Robinson-Schensted-Knuth (RSK) correspondence) 
also called tropical combinatorics. These involve properties of the $GL(N,R)$ Whittaker functions, which are generalizations
of Bessel functions. Later, it was shown by Borodin et al. \cite{logboro} 
that this generating function takes the form of a Fredholm determinant. 
This form allowed them to perform an asymptotic analysis for long DP and to prove again convergence 
of the PDF of the free energy to the GUE Tracy-Widom distribution. 
Finally, the O Connel-Yor model of the semi-discrete polymer \cite{semidiscret1}, which leads to an exactly solvable
hierarchy, can be obtained as a limit of the log-gamma polymer \cite{logsep2}. 
It would be of great interest to extend the Bethe Ansatz replica method to 
the discrete models. Recently, it was discovered by Brunet \cite{Brunet1}  that eigenfunctions of the
replica transfer matrix of the log-gamma polymer on the square lattice
can be constructed using a lattice version of the Bethe ansatz.
The present paper aims at studying these eigenfunctions, and from them to
calculate the generating function for the integer moments $\overline{Z^n}$ of the partition sum of the
log-gamma polymer. Here we treat the case of fixed endpoints.
The generating function is found to take the form of a Fredholm determinant
for all polymer lengths. 

This goal may appear hopeless at first sight, since the integer moments
$\overline{Z^n}$ cease to exist for $n \geq \gamma$ where 
$\gamma$ is the parameter of the model and the exponent of
the power law fat tail. However, our generating function reproduces all
existing integer moments. Furthermore, it suggests an analytic
continuation, inspired from Mellin-Barnes identities, 
which leads us to a conjecture for the Laplace transform of $P(Z)$
in the form a Fredholm determinant, with an (analytically continued) 
kernel. We use it to obtain the asymptotic behavior of the PDF of
the free energy $\ln Z$ at large polymer lengths.
In the limit of a very long DP, it yields convergence to the GUE Tracy-Widom distribution
up to non-trivial average and variance that we calculate. 
Our asymptotic predictions coincide with the result of Borodin et al. \cite{logboro} 
obtained by completely different methods (using the formula obtained in \cite{logsep2}).
In addition, we obtain the dependence in the end-point position on the lattice, e.g. the exact elastic coefficient at large times. We perform some numerical
checks of these results.

A more ambitious goal is then to show that the kernel obtained here is
equivalent to the one obtained in Borodin et al. \cite{logboro}. Most
steps of the correspondence are achieved and detailed here. However, the last
step involves the use of heuristics, although we present some hints
that it is correct.

Of course, as we show, our results also reproduce the ones of the continuum model, both at the level of the
Bethe-Ansatz (the Lieb-Linger model) and of the final result, i.e. our kernel reproduces
the finite time kernel for the corresponding KPZ/DP continuum model \cite{we,dotsenko}. In yet another
limit it also provides a Bethe Ansatz solution to the semi-discrete polymer problem \cite{semidiscret1}.

In general, the present work opens the way to explore the connections between 
quantum integrability and tropical combinatorics.

The outline of the paper is as follows. In Section \ref{Model} we recall the log-Gamma DP problem introduced by Seppalainen and introduce some useful notations. In Section \ref{EvolutionEquation} we present the ansatz discovered by Brunet. In Section \ref{TimeEvolution} we detail how this ansatz can be used to recursively compute the integer moments $\overline{Z^n}$, in particular we identify the weighted scalar product that makes the Brunet states orthogonal and (presumably) complete.
In Section \ref{LLlimit} we identify a scaling limit that relates the continuum model to the discrete one studied here. In Section \ref{Norm} we conjecture a formula for the norm of the Brunet functions that generalizes the Gaudin formula. In Section \ref{LargeL} we show how the Bethe-Brunet equations are solved in the "thermodynamic" limit. This allows us to find in Section \ref{Moments} an explicit formula for $\overline{Z^n}$. In Section \ref{GF} we perform an analytic continuation leading to a conjecture for the Laplace transform of the PDF of $Z$, as well as a formula for the PDF at fixed length. This is used in Section \ref{Asymptotic} to explicitly show the KPZ universality class and convergence of the fluctuations of $\log Z$ to the Tracy-Widom GUE distribution. In Section \ref{Comparison} we compare our results to those obtained in \cite{logboro}. Section \ref{Concl} summarizes the main conclusions of the paper, and a series of Appendices present some conceptual discussions and technical details.

\section{Model}\label{Model}

\subsection{Model}

\begin{figure}[h]
   \centering
   \includegraphics[scale=1]{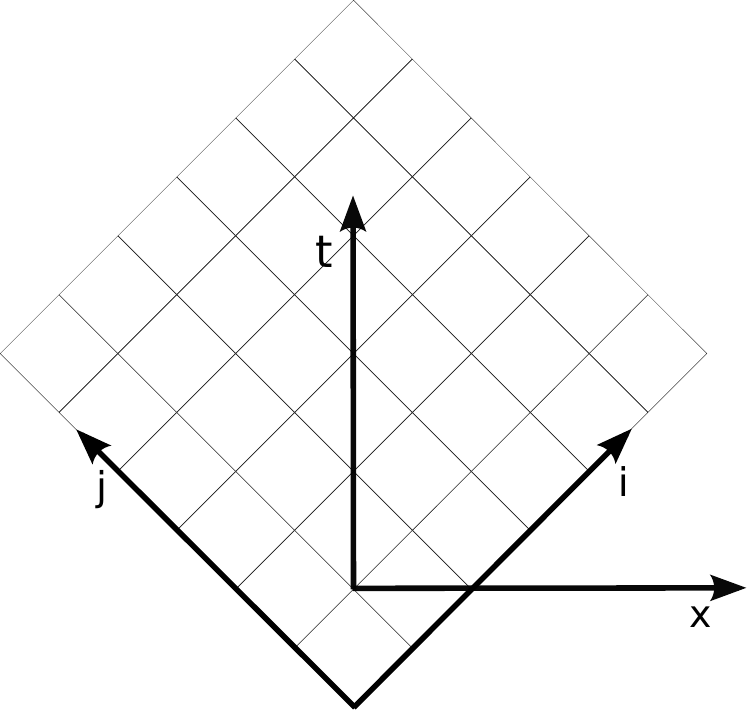}
   \caption{The two coordinate systems for the square lattice, see (\ref{coord12}). 
   The starting point of the path is $(i=1,j=1)$, which corresponds to the origin $(0,0)$ in
   the $(x,t)$ coordinates.
    }
   \label{coord}
\end{figure}

The log-Gamma directed polymer (DP) introduced by Seppalainen \cite{logsep1} is defined as follows. 
Consider the square lattice $(i,j) \in \mathbb{Z}^2$ and the set of directed up-right paths (directed
polymers) from $(1,1)$ to $(I,J)$. To emphasize the directed nature of the problem, 
we define $(x,t)$, with each coordinate running through one diagonal of the square lattice (see Fig. \ref{coord}): 
\be \label{coord12} 
 t =  i+j-2  \quad , \quad x = \frac{i-j}{2} 
\ee
so that the $x$ (space) coordinate of the points on a line with $t$ (time) even  (resp. odd) are integers (resp. half integers).
With this definition a directed path contains only jumps from $(x,t)$ to $(x + \frac{1}{2} , t+1) $ or $(x - \frac{1}{2} , t+1) $.
We define $Z_{t}(x)$ the (finite temperature) partition sum of the directed paths from $(0,0)$ to $(x,t)$:
\be \label{ps1} 
Z_{t}(x) = \sum_{ \pi : (0,0) \to (x,t) } \prod_{ (x',t') \in \pi } w_{x',t'}
\ee
in terms of the Boltzmann weights $w_{x,t} = e^{ - V_{x,t}}$ defined on the site of the lattice (the temperature is set to unity). In the
simplest (i.e. homogeneous) version of the log-Gamma DP model the $w_{x,t}$ are i.i.d. random variables distributed according to the inverse-Gamma distribution:
\begin{equation}
P(w) dw = \frac{1}{\Gamma(\gamma)} w^{-1 - \gamma} e^{-1/w} dw
\end{equation}
with parameter $\gamma >0 $. In the following $\overline{(.)}$ denotes the average over $w_{x,t}$ ("disorder average").

Our goal is to calculate the PDF of (minus) the free energy, $\ln Z_{t}(x)$, equivalently $P(Z_{t}(x))$. In the spirit of the recent
works on the replica Bethe Ansatz approach to the continuum directed polymer, we start by calculating the 
integer moments $\overline{ Z_{t}(x)^n} $ with $n \in \mathbb{N}$. Clearly these moments  do not exist for
$n \geq \gamma$, as can be seen already \footnote{$Z_{t}(x)$ always contains the statistically independent factors $w_{0,0}$ and $w_{x,t}$, corresponding to the endpoints.}
from the one-site problem $Z_0(0)=w_{0,0}=w$ whose moments are:
\begin{equation}\label{mo}
\overline{w^n} = \frac{ \Gamma( \gamma - n) }{\Gamma(\gamma)} 
\end{equation}
for $n<\gamma$, and diverge for $n \geq \gamma$. This makes a priori the problem of the
log-Gamma polymer more difficult to study using replica. However, note that 
(\ref{mo}) is valid more generally for $Re(n)< \gamma$ and possesses a simple analytic continuation
to the complex $n$ plane (minus the poles) via the $\Gamma$ function as given in (\ref{mo}). For this example,
and for more general ones, we show in \ref{appanalytic} how to obtain the Laplace transform $\overline{e^{- u w}}$ from
the integer moments (\ref{mo}).

This gives some hope to calculate the Laplace transform of $P(Z_{t}(x))$ with the sole knowledge of
its integer moments, via an analytic continuation, in the spirit of \ref{appanalytic}. 
The moment problem was a challenge for the
case of the continuum directed polymer due to the too rapid growth 
of the moments $\overline{Z^n} \sim e^{n^3 t}$. Here, the difficulty 
is the existence of poles in the moments, however
the situation for the analytic continuation appears more favorable.

\subsection{Rescaled Potential}

From now on we restrict ourselves to $\gamma>1$ and for convenience we normalize the weights so that 
their first moment is unity. We thus define:
\begin{eqnarray}
&& w =  \frac{1}{\gamma -1} \tilde w  = e^{-V} \nonumber \quad , \quad 
V = \tilde V + V_0  \quad , \quad e^{- V_0} = \frac{1}{ \gamma -1}
\end{eqnarray}
such that the integer moments become: 
\begin{equation} \label{hn}
h_{n} = \overline{ e^{ -n \tilde V} } =\frac{(\gamma -1)^n}{(\gamma -1) ... (\gamma-n)}  = \prod_{ k = 0} ^{ n-1} \frac{4}{ 4- k \bar{c} }
\end{equation}
where we introduced the interaction parameter:
\begin{equation}
\bar{c} = \frac{4}{ \gamma -1}  >0 .
\end{equation}
In particular, $h_0 = h_1 =1$. 


\section{Evolution equation and Brunet Bethe ansatz}\label{EvolutionEquation} 

\subsection{Evolution equation}

The partition sum of the directed polymer defined by (\ref{ps1}) 
can be calculated recursively as:
\begin{equation}\label{evolutionequation}
Z_{t+1}(x) = e^{- V_{x,t}}  \left(Z_t(x-\frac{1}{2}) + Z_t(x+\frac{1}{2}) \right) \quad , \quad Z_{0}(x) = e^{ - V_{0,0}} \delta_{x,0}
\end{equation}

The moments of the partition sum are conveniently encoded in the "wavefunction" $\psi$, defined on $\mathbb{Z}^n$ (for $t$ even) 
and $(\mathbb{Z}+ \frac{1}{2})^n$ (for $t$ odd)  as
\begin{equation}
\psi_t(x_1,...,x_n) = 2^{-nt} e^{V_0 n (t+1)}  \overline{Z_t(x_1) \cdots Z_t(x_n)} 
\end{equation}
which satisfies the evolution equation
\begin{eqnarray}\label{evolpsi}
\psi_{t+1}(x_1,\cdots,x_n) & = & \frac{1}{2^n} a_{x_1 , \cdots ,x_n}  \sum_{(\delta_1,\cdots,\delta_n) \in \{-\frac{1}{2},\frac{1}{2}\}^n} \psi_t(x_1 - \delta_1,\cdots,x_n - \delta_n) \nonumber \\
\end{eqnarray}
where we note:
\begin{equation} \label{defa} 
a_{x_1 , \cdots,x_n} =  \overline{  e^{ -\sum_{\alpha=1}^n \tilde V_{x_\alpha,t+1} } }  = \prod_{x} h_{ \sum_{\alpha=1}^n \delta_{x,x_\alpha}}
\end{equation}
and $h_n$ defined as in (\ref{hn}).

\subsection{ Bethe-Brunet Ansatz}

Consider the eigenvalue problem:
\bea \label{eigeneq} 
\psi_{\mu}(x_1,\cdots,x_n) & = &  \theta_\mu  ~ \frac{1}{2^n} a_{x_1 ,\cdots, x_n} \sum_{(\delta_1,\cdots,\delta_n) \in \{-\frac{1}{2},\frac{1}{2}\}^n} \psi_\mu(x_1 - \delta_1,\cdots,x_n - \delta_n)
\eea 
It was found by Brunet \cite{Brunet1} that fully-symmetric solutions $\psi_\mu$ 
of (\ref{eigeneq}) can be obtained as superpositions of plane waves in a form that generalizes the usual Bethe Ansatz:
\begin{equation}\label{brunetansatz}
\psi_\mu(x_1,\cdots,x_n) = \sum_{\sigma \in S_n} A_\sigma \prod_{\alpha=1}^n z_{\sigma(\alpha)}^{x_\alpha}  \hspace{0.3 cm},\hspace{0.3 cm}  A_\sigma = \prod_{1 \leq \alpha < \beta \leq n } (1+\frac{\bar{c}}{2} \frac{sgn(x_\beta-x_\alpha + 0^+)}{t_{\sigma(\alpha)} - t_{\sigma(\beta)}} ) 
\end{equation}
with
\begin{equation} \label{corr}
z_\alpha = e^{ i \lambda_\alpha}   \hspace{0.3 cm},\hspace{0.3 cm} t_\alpha = i \tan (\frac{\lambda_\alpha}{2}) =\frac{z_\alpha-1}{z_\alpha+1} 
\end{equation}
These solutions $\psi_\mu$ are parametrized by a set of (distinct) complex variables $\{ z_1,\cdots,z_n \}$.
It is convenient to parametrize the $z_\alpha$ in terms of variables $\lambda_\alpha$ as above, with
$ -\pi < Re(\lambda_\alpha) \leq \pi$, which we call rapidities by analogy with the continuum case
(see discussion below). The eigenvalue associated with $\psi_\mu$ is then given by:
 \footnote{the first factor $\prod_{\alpha=1}^{n} z_\alpha^{\frac{1}{2}}$ was absent in Brunet's formula due to a different choice
of coordinates $x'=x+t/2$.} 
\begin{equation}\label{thetamu}
\theta_\mu = \prod_{i=1}^{n} z_\alpha^{\frac{1}{2}} \frac{1+ z_\alpha^{-1}}{2} 
\end{equation}
The property (\ref{eigeneq}) is easily checked for all $x_\alpha$ distinct, in which case it is similar to the continuum case
\cite{ll,gk}. The case
where there are two coinciding $x_\alpha$ is reminiscent of the matching condition of the continuum case. Verifying
the property (\ref{eigeneq}) for an arbitrary number of coinciding points is non-trivial, and is found to work {\it only}
when the $h_n$ in (\ref{defa}) have values precisely given by (\ref{hn}) \cite{Brunet1}. Hence this integrability
property is a special property of the inverse Gamma distribution \footnote{there are other solvable cases, by different methods, such as zero temperature
model of \cite{Johansson2000}, solved in terms of a determinantal process related to free fermions.}. 
Until now the possible values of the $z_\alpha$ remain unspecified. As an intermediate stage in 
our calculation we impose here for convenience 
periodic boundary conditions $\psi(x_1,\cdots , x_\alpha +L ,\cdots,x_n) = \psi(x_1,\cdots,x_n)$, $\alpha=1,\cdots,n$,
i.e. a system of finite number of sites $L$. This can be satisfied if the rapidities satisfy the 
generalized Bethe equation \cite{Brunet1}:
\begin{equation} \label{BAE}
e^{i \lambda_{\alpha}L} = \prod_{1 \leq \beta \leq n, \beta \neq \alpha} \frac{2 t_\alpha- 2 t_\beta+\bar{c}}{2 t_\alpha-2 t_\beta-\bar{c}} =  \prod_{1 \leq \beta \leq n, \beta \neq \alpha} 
 \frac{2 \tan (\frac{\lambda_\alpha}{2})-2 \tan (\frac{\lambda_\beta}{2})- i \bar{c} }{2 \tan ( \frac{\lambda_\alpha}{2})-2 \tan (\frac{\lambda_\beta}{2}) + i \bar{c}} 
\end{equation}
for $\alpha=1,\cdots,n$, which are derived exactly as in the continuum case.

\section{Time evolution of the moments, symmetric transfer matrix}\label{TimeEvolution}

\subsection{Symmetric transfer matrix and scalar product}

In this section we motivate the introduction of a peculiar weighted scalar product, for which the Brunet functions form an orthogonal set.  The Brunet functions diagonalize the evolution equation~(\ref{evolpsi}), which is not encoded by a symmetric transfer operator since
the variable $a_{x_1,\cdots,x_n}$ depends only on the arrival point. This can be traced to the recursion (\ref{evolutionequation}), which 
counts the contribution of the disorder only at the points on the line at $t+1$. Hence the Brunet functions have no reason to form an orthogonal set for the canonical scalar product, and we indeed find that they do not. On the other hand, if we consider the change of function $\tilde{\psi}(x_1,\cdots,x_n) = \frac{1}{\sqrt{a_{x_1,\cdots,x_n}} } \psi(x_1,\cdots,x_n) $, (\ref{evolpsi}) now reads
\bea \label{evolsym}
\!\!\!\!\!\!\!\!\!\!\!\!\!\!\!\!\!\!\!\!\!\!\!\! \tilde{\psi}_{t+1}(x_1,\cdots,x_n) = \sqrt{a_{x_1,\cdots,x_n}}  \sum_{(\delta_1,..\delta_n) \in \{-\frac{1}{2},\frac{1}{2}\}^n} \sqrt{a_{x_1 - \delta_1 ,\cdots,x_n -\delta_n}}  \tilde{\psi}_t(x_1 - \delta_1,\cdots,x_n - \delta_n)
\eea
The disorder now appears in a symmetric way, and the transformed Brunet functions $\tilde{\psi}_{\mu}$ naturally appear as eigenvectors of an Hermitian transfer operator, with the same eigenvalue $\theta_{\mu}$ as before. This shows that $\theta_{\mu} \in \mathbb{R}$ . Since  (\ref{evolsym}) involves the evaluation of a function both at integer coordinates and half-odd integer coordinates, this operator acts on the function defined on $\mathbb{Z}^n \oplus (\mathbb{Z} + \frac{1}{2} )^n $.  It appears more convenient to consider the evolution equation that links $t$ and $t+2$: this defines the transfer matrix $T_n$:
\bea \label{transfermatdef}
\tilde{\psi}_{t+2}  = T_n \tilde{\psi}_t
\eea
which is thus naturally defined as an Hermitian operator on $L^2(\mathbb{Z}^n )$, and for which the Brunet states $\tilde{\psi}_{\mu}$  are eigenvectors with eigenvalues $e^{-2 E_{\mu} } = \theta_{\mu} ^2 >0 $
\begin{equation} \label{eigen}
\theta_{\mu} ^2 = e^{-2 E_{\mu}} =  \prod_{\alpha=1}^{n} \frac{ z_\alpha +2 +  z_\alpha^{-1}}{4} = \prod_{\alpha=1}^{n} \frac{1}{1-t_\alpha^2} 
\end{equation}
where the last equation is an equivalent form, using that $z_\alpha=(1+t_\alpha)/(1-t_\alpha)$.

To be more precise, we have chosen to work with periodic boundary conditions and we thus consider $T_n$ as an operator that acts on the function defined on $\{0,\cdots, L-1\}^n$, which has dimension $L^n$. This is only a convenient choice and should have no effect on the results 
for the case of interest here, i.e. a polymer with a fixed starting point,
as long as we consider $t<L$: in this case the polymer does not ever feel the boundary. In the end we will consider the limit
$L \to \infty$ at fixed $t$, so that the polymer never feels the boundary. 

Going back to the original wavefunctions, the above construction partially justifies the claim that the original Brunet states $\{ \psi_{\mu} \}$ 
given in (\ref{brunetansatz}) 
form a {\it complete} basis of the symmetric functions on $\{0,\cdots, L-1\}^n$, and that it is {\it orthogonal} with respect to the following weighted scalar product
\begin{equation}\label{wps}
\langle \phi , \psi \rangle = \sum_{ (x_1 , \cdots , x_n ) \in  \{0,\cdots, L-1\}^n } \frac{1}{a_{x_1,\cdots,x_n}} \phi^*(x_1,\cdots,x_n)  \psi(x_1,\cdots,x_n)
\end{equation}
We have not attempted to provide a general proof of this statement (a usually challenging goal
when dealing with Bethe Ansatz), however we did explicitly check it for various low values of $(L,n)$.
We will thus proceed by assuming that it is correct. 

\paragraph{}
We conclude this section with a minor remark on a special case: if there is a solution of the Brunet equation with $z_i = -1$,  then $e^{-2 E_{\mu} } = 0$ and the Brunet state is ill-defined. In fact, it is easy to see that $T \psi_{\mu}  =0$ if and only if $M \psi_{\mu}  =0 $ with $M$ the transfer matrix without disorder, which can be diagonalized using plane waves. Hence to have a well-defined complete basis, one has to complete the Brunet states with the symmetric plane waves with vanishing eigenvalues that exist when $L$ is even. 
These additional states do not play any role in the following (since they
correspond to zero eigenvalues) but they 
are important to assess the validity of the completeness property.

\subsection{Time-evolution of the moments}

This formalism allows us to give a simple expression for the moments with arbitrary endpoints:
\be \label{evo1} 
\overline{ Z_{t} (x_1) \cdots Z_{t} (x_n) }= 2^{nt} \left( \frac{ \bar{c} }{4} \right)^{n(t+1)}  \psi_t( x_1,\cdots,x_n)
\ee
Since the Brunet states form a complete basis of the symmetric functions on $\{0,\cdots, L-1\}^n$, which are orthogonal with respect to the scalar product (\ref{wps}) and since the initial condition
\be
\psi_0(x_1,\cdots,x_n) = h_n \prod_{\alpha=1}^n \delta_{x_\alpha , 0}
\ee
is symmetric in position space, one can write the decomposition of the initial condition on the
Brunet-Bethe states as:
\begin{equation}
\psi_0 = \sum_{ \mu} \frac{ \langle \psi_\mu , \psi_0  \rangle }{ || \psi_{\mu} || ^2} \psi_\mu = \sum_{ \mu} \frac{ n! }{ || \psi_\mu || ^2} \psi_\mu 
\end{equation}
 using the explicit expression (\ref{brunetansatz}) for the (un-normalized) eigenstates. The simple iteration of the evolution equation (\ref{evolpsi}) directly leads to, for all $t \in \mathbb{N}$:
\begin{equation}
\psi_{t} = \sum_{ \mu} \frac{ n! }{ || \psi_\mu || ^2}  (\theta_{\mu})^t \psi_\mu 
\end{equation}
and thus
\begin{equation}
\overline{ Z_{t} (x_1) \cdots Z_{t} (x_n) } = 2^{nt  } \left( \frac{ \bar{c} }{4} \right)^{n(t+1)} 
\sum_{ \mu} \frac{ n! }{ || \psi_\mu || ^2} (\theta_{\mu})^t  \psi_\mu(x_1,\cdots ,x_n)
\end{equation}
Using that:
\be
\psi_\mu( x ,...,x) = n! \left( \prod_{\alpha=1}^nz_{\alpha} \right)^x
\ee
for any eigenstate $\mu$ given by (\ref{brunetansatz}), we finally obtain
the integer moment of the DP with fixed starting point at $(0,0)$ and endpoint 
at $(x,t)$ as:
 \begin{equation}\label{momentn}
\overline{ Z_{t} (x)^n } = 2^{nt  } \left( \frac{ \bar{c} }{4} \right)^{n(t+1)} \sum_{ \mu} \frac{ (n!)^2 }{ || \psi_\mu || ^2}  (\theta_{\mu})^{t}   \left( \prod_{\alpha=1}^nz_{\alpha} \right)^x
\end{equation}
where we recall $\theta_\mu$ to be given by (\ref{thetamu}). 
Hence the only remaining unknown quantities here are the norm of the Brunet states, 
and we will now calculate them in the infinite size limit $L \to \infty$.

Before we do so, let us indicate how the present discrete model recovers the
continuum model in some limit, in particular how the discrete space-time quantum mechanics
recovers the standard continuum one.

\section{The continuum/Lieb-Liniger limit}\label{LLlimit}

It is interesting to note that the Brunet equations (\ref{BAE}) and the form of the eigenfunctions (\ref{brunetansatz})
tend to those of the Lieb-Liniger model (LL) as given by the standard Bethe ansatz solution if one takes
the limit of small $\lambda_i$ and $\bar c$ simultaneously. In such limit, one has $t_i \simeq i \frac{\lambda_i}{2}$.

\bigskip
More precisely, to understand the correspondence between the continuum LL model \cite{ll} and the present discrete model, we must reintroduce a lattice spacing ${\sf a}$ that sets the dimension of the parameters of the continuum case. We define
\bea \label{LLscaling}
&& \lambda_\alpha = {\sf a} \lambda_\alpha^{LL}  \quad , \quad \bar c = {\sf a} \bar c^{LL}  \quad , \quad x_\alpha = \frac{x_\alpha^{LL}}{{\sf a}} \quad , \quad t = \eta \frac{t^{LL}}{{\sf a}^2} 
\eea 
where we keep temporarily $\eta$ as a free parameter. At finite size we must also define the periodicity of the LL model, $L^{LL}=${\sf a}$ L$.

If one now takes the LL limit defined by ${
\sf a} \to 0$ with the quantities of the continuum (labelled $LL$) fixed, one recovers 
from (\ref{brunetansatz})-(\ref{corr}) the usual Bethe wavefunctions for the LL model, with rapidities $\lambda_\alpha^{LL}$ and
(attractive) interaction parameter $c^{LL} = - \bar c^{LL}<0$. From (\ref{BAE}) we also recover the usual Bethe equations for the LL model:
\begin{equation} \label{BELL} 
 e^{ i \lambda_\alpha^{LL} L^{LL}}  =  \prod_{\beta \neq \alpha} \frac{\lambda_\alpha^{LL}- \lambda_\beta^{LL} - i\bar{c}^{LL}}{\lambda_\alpha^{LL}- \lambda_\beta^{LL}+ i \bar{c}^{LL}}
\end{equation}
The parameter $\eta$ tunes the correspondence between the LL time and our discrete time $t$: in the $LL$ case the time-evolution of an eigenfunction $\mu$ is encoded through the multiplication by a factor $e^{ -E^{LL}_{\mu} t^{LL}} = e^{ - \sum_{i=1}^n (\lambda_i^{LL})^2 t^{LL} }$, which should be equal to the LL limit of $ (\theta_{\mu} )^{t} $. This implies
\bea \label{energyexp} 
t^{LL} \sum_{\alpha=1}^n (\lambda_\alpha^{LL})^2 = - \lim_{{\sf a}\to 0} \eta \frac{t^{LL}}{{\sf a}^2}   \sum_{\alpha=1}^n \log\left( \frac{ e^{  \frac{ i {\sf a} \lambda_{\alpha}^{LL} } {2} } + e^{ - \frac{ i{\sf a} \lambda_{\alpha}^{LL} } {2} }  }{2}   \right) = \eta t^{LL} \sum_{\alpha=1}^n \frac{(\lambda_\alpha^{LL})^2}{8}
\eea
If we now follow standard conventions and definitions of the LL model, see e.g. \cite{we,we-flat}, this 
implicates $\eta = 8$. With this choice, the time-evolution of our wavefunction is consistent with the one of the continuum model.

To further extend the correspondence to the moments of the partition sum, we must compare the
formula (\ref{evo1}) with the similar evolution for the LL model (where the wavefunction was simply equal to the
moment). The correspondence thus reads:
\bea \label{LLcorresponence}
&& \overline{ Z^{LL}_{t^{LL}} (x_1^{LL}) \cdots Z^{LL}_{t^{LL}} (x_n^{LL})  }^{V_{LL}} = \lim_{ {\sf a} \to 0} 2^{- nt} \left( \frac{4 }{ \bar{c}} \right)^{n(t+1)} \overline{ Z_{t} (x_1) \cdots Z_{t} (x_n)  }^{w} \\
&& Z^{LL}_{t^{LL}} (x^{LL})   \equiv_{\rm in law}  \lim_{ {\sf a} \to 0} 2^{- t} \left( \frac{4 }{ \bar{c}} \right)^{(t+1)} Z_{t} (x) 
\eea
where on the right the limit has to be taken using (\ref{LLscaling}). We have emphasized that averages in 
the continuum model ($LL$) are computed for a Gaussian potential $V_{LL}$, which is distinct from the quenched disorder in the discrete model. The second equation
states the equivalence "in law"  between the discrete log-gamma DP model
in the small lattice spacing limit, and the continuum DP model~\footnote{strictly, 
this could be considered as a conjecture since both models have an ill-defined moment problem (see however below).}. For a precise
definition of the continuum DP model, including $V_{LL}$, with
the same conventions, see e.g. \cite{we,we-flat}. 

Note that we have somewhat "reverse-engineered" here, since
one can also establish (\ref{LLcorresponence}) by directly starting from the evolution equation for the moments (\ref{evolpsi}), without any knowledge of the Bethe ansatz solution. A similar calculation was performed in \cite{brunet}.
The present considerations thus provide a useful consistency check. Note that the various continuum limits are also discussed in \cite{BorodinMacdo}, Section 5.
 
In the following, we note $\simeq_{LL} $ the $LL$ limit, which is the limit of small ${\sf a}$ with the scaling (\ref{LLscaling}).
Note that it corresponds to the limit of $\gamma = 1 + 4/({\sf a} \bar c_{LL}) \to \infty$ in the log-gamma DP model. 

\section{Norm of the eigenstates}\label{Norm} 

Here we will guess a general formula for the norm of the eigenstates for the discrete model (the Brunet states). The approach involves some heuristics, but the final formula reproduces all numerical verifications that we performed for small values of $n$, as it is summarized in \ref{appgaudin}. The complete proof of the formula will surely be involved, e.g. as it was the case in the continuum case~\cite{gk}.

Let us recall the formula for the norm for the LL model (with periodic boundary conditions):
\begin{equation}\label{norme1LL}
||\mu||_{LL}^2 = n!  \prod_{1 \leq \alpha < \beta \leq n }  \frac{   (\lambda_{\alpha}^{LL} - \lambda_{\beta}^{LL}  ) ^2 + (\bar{c}^{LL} )^2  }   {    (\lambda_{\alpha}^{LL}  - \lambda_{\beta}^{LL}  ) ^2} \det{ G^{LL}  }
\end{equation}
where $G^{LL}$ is the Gaudin matrix whose entries are:
 \bea
&& G_{\alpha \beta}^{LL} = \delta_{\alpha  \beta} \left( L + \sum_{\gamma=1}^n K (\lambda_{\alpha}^{LL} - \lambda_{\gamma}^{LL}) \right) - K ( \lambda_{\alpha}^{LL} - \lambda_{\beta}^{LL}) \\
&& K(x) =  \frac{- 2 \bar c^{LL}}{x^2 + (\bar c^{LL})^2} 
\eea
A useful remark is that the entries of the Gaudin matrix in the LL case 
are the derivatives of the logarithm of the LL Bethe equations (\ref{BELL}). 

Let us {\it assume} that this property still holds. From the Brunet-Bethe equations
(\ref{BAE}) we can then summarize that in the present case:
\begin{equation}
G_{\alpha \beta} = \frac{1}{i} \frac{\partial}{ \partial \lambda_\beta} \left( \log \left( e^{i \lambda_\alpha L} \prod_{j\neq i} \frac{2 t_\alpha- 2 t_\beta-\bar{c}}{2 t_\alpha-2 t_\beta+\bar{c}} \right) \right)
\end{equation}
Using that $\partial_{i \lambda_\alpha} t_\alpha = \frac{1-t_\alpha^2}{2}$,
this leads to a modified Gaudin matrix:
 \begin{equation}\label{modifiedgaudin}
G_{\alpha \beta} = \delta_{\alpha  \beta} \left( L + (1 - t_{\alpha}^2) \sum_{\gamma=1}^n 
\tilde K (t_{\alpha} - t_{\gamma}) \right) - (1- t_{\beta}^2)  \tilde K ( t_{\alpha} - t_{\beta}) 
\end{equation}
with
\begin{equation}
\tilde K (t)  = \frac{- 2 \bar{c} }{ - 4 t^2 + \bar{c}^2} 
\end{equation}

And our final conjecture for the norm is:
\begin{equation}\label{norme1}
||\mu||^2 = n!  \prod_{1 \leq \alpha < \beta \leq n }  \frac{   (2 t_{\alpha} -2  t_{\beta} ) ^2 - \bar{c}^2  }   {   (2 t_{\alpha} - 2 t_{\beta} ) ^2  } \det{ G }
\end{equation}
where the $t_\alpha$ are given by (\ref{corr}) and are solutions of the
Bethe-Brunet equations (\ref{BAE}). This formula is constructed to coincide with the formula (\ref{norme1LL}) in the $LL$ limit. It is remarkable, since it could have been constructed without knowing the definition (\ref{wps}) of our peculiar weighted scalar 
product, and as such it is another manifestation of the nice properties of integrable systems. 

We will now proceed assuming this formula to be correct, and later on the
way we will indeed carry more indirect checks of its validity.

\section{Large $L$ limit}\label{LargeL}

In this section we obtain the string eigenstates in the large $L$ limit, as well as
expressions for their eigenvalue (energy), momentum, phase-space contribution and norm.

\subsection{Strings}

We now turn to the large $L$ limit where the analysis can be made more precise, and the Bethe-Brunet equations (BBE) 
can be solved in an
asymptotic sense, the crucial point being the existence of string-states. Let us analyze the BBE equations (\ref{BAE}) in the large $L$ limit:
\begin{equation} \label{BAE2} 
e^{ i \lambda_\alpha L} = \prod_{\beta \neq \alpha} \frac{2 t_\alpha- 2 t_\beta+\bar{c}}{2 t_\alpha-2 t_\beta-\bar{c}}
\end{equation}
where we recall  $t_\alpha = i \tan (\frac{\lambda_\alpha}{2}) $.
The analysis parallels the one of the continuum problem, with a few (important) differences. 

If all the $\lambda_\alpha$ are real, we note $\lambda_\alpha = \hat k_\alpha \in \mathbb{R} $
and the $t_\alpha$ are pure imaginary numbers, $t_\alpha = i \frac{k_\alpha}{2}$ with $k_\alpha \in \mathbb{R} $.
This situation is very similar to the LL model: the left hand side in (\ref{BAE2}) is
$e^{ i \hat k_\alpha L}$ and the quantization of the variables 
$\hat k_\alpha$ is similar to the free momenta quantization, plus 
corrections of order $O(1/L)$. The momentum variable $\hat k_\alpha$
belongs to the first Brillouin zone, $]-\pi,\pi]$, which is natural since 
we are studying a discrete model. This situation corresponds to $1$-strings, also called particles. 
Note that $k_\alpha = 2 \tan(\hat k_\alpha/2)$, and the two quantities become identical
only in the LL limit, where both are small (see below). 

If however one of the $\lambda_\alpha$ has an imaginary part $\delta$, which we assume to be positive,
the left hand side of the equation tends to zero exponentially as $e^{- \delta L} $.
This indicates that there must exist another $t_\beta$ such that
\begin{equation}
t_\beta = t_\alpha + \frac{\bar{c}}{2} + O( e^{ - \delta L} )
\end{equation}
or equivalently 
\begin{equation}
\tan( \frac{\lambda_\beta}{2} ) = \tan( \frac{ \lambda_\alpha }{2} ) - i \frac { \bar{c} }{2}  + O( e^{ - \delta L} ) 
\end{equation}
Since $z \to \tan(z)$ preserves the sign of the imaginary part, we get a new eigenvalue with a lower imaginary part and we can continue the procedure. If the imaginary part of $t_\gamma$ is negative we get that there must exist $\gamma'$ such that $t_{\gamma'} = t_\gamma - \frac{\bar{c}}{2} + O( e^{ - \delta L} )$, and this procedure has to terminates at some point. 
In fact, as in the Lieb-Liniger case, we believe that it is a general fact that each set of $i t_\alpha$ solution to the Brunet equations is self-conjugate, and that in the large-time limit the $t_\alpha$ organize themselves as depicted above.

\bigskip

To conclude, the key idea is that in the large $L$ limit, a set $\{t_\alpha\}$ that solves the Brunet equations is divided into strings such that inside each string the $t_\alpha$ are distant from each other by $\frac{\bar{c}}{2} $. A general eigenstate is given by partitioning $n$ into $n_s$ strings, each string containing $m_j$ particles where the index $j=1,\cdots,n_s$ labels the string.
We can thus write all the $t_\alpha$, $\alpha=1,\cdots,n$, in the form:
\begin{equation} \label{string} 
t_\alpha = t_{j,a} = i \frac{ k_j}{2} + \frac{\bar{c}}{4} ( m_j + 1 - 2a ) +  \frac{\delta_{j,a}}{2} 
\end{equation}
where we introduce an index $a=1,\cdots, m_j$ that labels the rapidity inside a string, and 
$\delta_{j,a}$ are deviations that fall off exponentially with $L$. Hence inside the 
$j^{th}$ string the $t$ variables have the same imaginary part that is denoted by $\frac{k_j}{2}$.

 One easily sees that the strings of the present model reproduce the LL strings in the LL limit. For infinite
$L$ the correspondence reads:
\be \label{corr1} 
 t_\alpha = t_{j,a}  \simeq_{LL}  {\sf a} \lambda^{LL}_{j,a} + O({\sf a}^3) \quad , \quad  \lambda^{LL}_{j,a}  = i \frac{ k^{LL}_j}{2} + \frac{\bar{c}^{LL}}{4} ( m_j + 1 - 2a ) 
\ee
and the variables $k_j$ in (\ref{string}) correspond to leading order to the LL string momenta through the scaling $k_j \simeq {\sf a} k_j^{LL} +O({\sf a}^3)$.

%
%
%

{\it Restriction on the multiplicity of the string: } there is however an important difference with the case of LL strings. 
One can see that the mapping between $\lambda_\alpha$ and $t_\alpha$ is a bijection if  $|Re(t_\alpha)| < 1$, i.e. if $ \bar{c} < \frac{4}{m-1} $. Since $m\leq n$ this implies $ \bar{c} < \frac{4}{n-1} $ or equivalently $n < \gamma$, which is exactly the condition for the moment problem to be well-defined. In the LL limit we have $\gamma \to \infty$ and one recovers that there are no restriction on $m,n$.

\subsection{Eigenvalue of a string: energy}

Inserting (\ref{string}) into (\ref{eigen}) easily gives that the eigenvalue associated to a string state takes the form of a product:
\be
\theta_{\mu} = \prod_{j=1}^{n_s}  \theta_{m_j,k_j} 
\ee
where the contribution of a single string can be written in several forms 
\footnote{note that from (\ref{string}) $1-t_a=(1+t_{m+1-a})^*$ and complex conjugation amounts to change $k \to - k$.} 
\begin{eqnarray}\label{stringenergy}
 \theta_{m_j,k_j}  = && \left( \prod_{a=1}^{m_j}\frac{1}{ 1-t_{j,a}^2} \right)^{ \frac{1}{2}}  =  \left( \frac{2}{ \bar{c}} \right)^{ m_j}  \left(\frac{1}{ \left(\frac{-m_j \bar{c}+\bar{c}-2 i k_j+4}{2 \bar{c}}\right)_{m_j} \left(\frac{-m_j \bar{c}+\bar{c}+2 i  k_j+4}{2 \bar{c}}\right)_{m_j} } \right)^{\frac{1}{2}}\\
&& = \left(\frac{2}{ \bar{c}} \right)^{ m_j}  \left(\frac{  \Gamma(-\frac{m_j}{2} + \frac{ \gamma}{2} - i\frac{k_j}{\bar{c} } ) \Gamma(-\frac{m_j}{2} + \frac{ \gamma}{2} + i\frac{k_j}{\bar{c} })}{  \Gamma(\frac{m_j}{2} + \frac{ \gamma}{2} -i\frac{k_j}{\bar{c} } ) \Gamma(\frac{m_j}{2} + \frac{ \gamma}{2} + i\frac{k_j}{\bar{c} }) } \right)^{\frac{1}{2}}
\end{eqnarray}
which are equivalent for integer $m$. Here $(a)_m = a (a+1) ..(a+m-1)=\Gamma(a+m)/\Gamma(a)$ is the Pochhammer symbol and we reintroduced $\gamma = 1 + \frac{4}{\bar{c}} $ in the last expression.

Writing $\theta_{\mu} = e^{- E_{\mu} } $, one can verify the Lieb-Liniger limit:
\bea
E_{\mu} t \simeq_{LL}  \sum_{i=1}^{m_j} \left( m_j (k^{LL}_j)^2  - \frac{(\bar{c}^{LL})^2}{12} m_j (m_j^2-1) \right) t^{LL} .
\eea
in two ways. Either the easy way, on the starting expression (first equation in (\ref{stringenergy}) before summing over $a$) 
using (\ref{corr1}) and performing an expansion similar to (\ref{energyexp}). A more tedious way is to use the 
final expression in (\ref{stringenergy}) after summation over $a$. This is detailed in 
\ref{app:limits}, where the next higher order corrections $O({\sf a} ^2)$ are also given.

\subsection{Momentum of a string}

In the formula (\ref{momentn}) for $\overline{Z_t(x)}$, the temporal dependance appears through the eigenvalue whereas the position dependence appears through the factor $\left( \prod_{\alpha} z_\alpha \right)^x$ which also takes a simple form in string notations: $\prod_{\alpha} z_\alpha = \prod_{j=1}^{n_s} \prod_{a=1}^{m_j} \frac{1+t_{j,a}}{1-t_{j,a}}$, the contribution of a single string being

\begin{equation}\label{stringmomenta}
\prod_{a=1}^{m_j}  \frac{1+t_{j,a}}{1-t_{j,a}} =\frac{ \Gamma( -\frac{m_j}{2} + \frac{\gamma}{2} - i\frac{k_j}{\bar{c} }) \Gamma( \frac{m_j}{2} + \frac{\gamma}{2} +i\frac{k_j}{\bar{c} })}{\Gamma( \frac{m_j}{2} + \frac{\gamma}{2} - i\frac{k_j}{\bar{c} }) \Gamma( -\frac{m_j}{2} + \frac{\gamma}{2} + i\frac{k_j}{\bar{c} })} 
\end{equation}
As for the eigenvalue, one can check the Lieb-Liniger limit:
\begin{equation}
\left(\prod_{a=1}^{m_j}  \frac{1+t_{j,a}}{1-t_{j,a}} \right)^x \simeq_{LL} e^{ i m_j k_j^{LL} x^{LL}}
\end{equation}

\subsection{ Phase space}

The sum over all eigenstates in (\ref{momentn}) can be computed as follows:
as in the case of the Lieb-Liniger model \cite{cc-07}, regarding the quantization of its center of mass, each string state should be considered as a free particle in the large $L$ limit, with total momentum $K_j = \sum_{a=1}^{m_j} \lambda_{j_a} \in [- m_j \pi , m_j \pi] $ (we choose to restrict the momenta to belong to the first Brillouin zone, since we work on a discrete model). This property allows us to compute the Jacobian and therefore to express sums over Brunet eigenstates: we write
\begin{equation}\label{stringmomentum}
e^{ i L K_ j} =  \prod_{a=1}^{m_j } \frac{ 1+ t_{j,a} }{1- t_{j,a} } 
\end{equation}
where we effectively ignored the interaction with the other strings. We can thus rewrite the sum over string states
using (\ref{stringmomentum}) as: 
\begin{equation}
\sum_{ m_j string-states } \to \frac{L}{2 \pi} \int_{- m_j \pi} ^{ m_j \pi} dK_j \to \frac{L}{2 \pi} \int_{- \infty}^{ \infty}  dk_j \sum_{a=1}^{m_j} \frac{1}{ 1 - t_{j,a}^2}
\end{equation}
which, in comparison with the usual formula for the LL model $\frac{L}{2 \pi} m_j \int_{- \infty}^{ \infty}  dk_j$ has an additional
"Jacobian" factor.

\subsection{ Norm of the string states}

As in the Lieb-Liniger case, our analogous Gaudin-like formula for the norm (\ref{norme1}) 
has to be studied carefully in the limit of a large system size to obtain the formula
for the norm of the string states. The calculation is detailed in \ref{appgaudin2} and we only give here the result that the leading order in $L$ is

\bea \label{norme2}
\!\!\!\!\!\!\!\!\!\!\!\!\!\!\!\!\!\!\!\!\!\!\!\!\!\!\!\!\! ||\mu||^2 = n! L^{n_s}   \prod_{1\leq i < j  \leq n_s} \frac{4 (k_i-k_j)^2 + \bar{c}^2 (m_i + m_j)^2}{4(k_i-k_j)^2 + \bar{c}^2 (m_i - m_j)^2} \prod_{j=1}^{n_s} [ \frac{m_j}{ \bar{c}^{m_j-1} } (\sum_{a=1}^{m_j}  \frac{1}{1-t_{j,a}^2}) \prod_{b=1}^{m_j} (1-t_{j,b}^2) ]
\eea 
which is the generalization of the Calabrese-Caux formula in the case of the LL model 
\cite{cc-07}. The LL formula is recovered by setting all $t_j = 0$ in the above result. Note that it should be possible to derive a rigorous proof of this result and of the completeness of the Brunet states in the $L \to\infty$ limit, where one can use e.g. Plancherel type isomorphism techniques, as was done in \cite{BorodinQboson} for the $q$-Boson particle system.

\section{Formula for the integer moments $\overline{Z^n}$}\label{Moments}

We now have all the ingredients to compute the moments in the limit of
large system size $L \to \infty$ at fixed $t,x$. Using the results of the
previous section, (\ref{momentn}) can be rewritten
as:

\bea \label{momentn2}
&& \!\!\!\!\!\!\!\!\!\!\!\!\!\!\!\!\!\!\!\!\!\!\!\!\!\!\!\!\!\!\!\!\! \overline{Z_t(x)^n} = 2^{n t}  \left( \frac{ \bar{c} }{4} \right)^{n(t+1)}  n! \sum_{n_s=1}^n  \frac{1}{n_s!} \sum_{(m_1,..m_{n_s})_n} 
\prod_{j=1}^{n_s}  \int_{-
 \infty}^{+\infty} [ \frac{dk_j}{2 \pi} \sum_{a=1}^{m_j}  \frac{1}{1-t_{j,a}^2} ] 
\prod_{1\leq i < j  \leq n_s} \frac{4(k_i-k_j)^2 + \bar{c}^2 (m_i - m_j)^2}{4(k_i-k_j)^2 + \bar{c}^2 (m_i + m_j)^2} \nn \\
&&
\prod_{j=1}^{n_s} (\bar{c})^{m_j-1} \frac{1}{m_j (\sum_{a=1}^{m_j}  \frac{1}{1-t_{j,a}^2}) \prod_{b=1}^{m_j} (1-t_{j,b}^2)} 
\prod_{b=1}^{m_j} ( \frac{1}{1-t_{j,b}^2} )^{t/2} ( \frac{1+t_{j,b}}{1-t_{j,b}} )^{x} 
\eea 
where we have used that the sum over states can be written as $ \sum_{ \mu} = \sum_{n_s=1}^n  \frac{1}{n_s!} \sum_{(m_1,\cdots,m_{n_s})_n}  \sum_{ m_j string-states }  $, where $\sum_{(m_1,\cdots,m_{n_s})_n} $ means that we sum over all $n_s$-uplets  $(m_1 , \cdots, m_{n_s})$ such that $ \sum_{i=1}^{n_s} m_i  = n$, and the $n_s!$ factor avoids counting the same string state twice. Note the cancellation in that formula between the phase space Jacobian factor and a similar factor in the norm. The rescaling $k_i \to \bar{c} k_i $ and the use of the formula for the energy term (\ref{stringenergy}) and for the momentum term (\ref{stringmomentum}) directly gives our main formula
for the integer moments:
\bea \label{momentn3}
&&\!\!\!\!\!\!\!\!\!\!\!\!\!\!\!\!\!\!  \overline{Z_t(x)^n} =  n!  \sum_{n_s=1}^n  \frac{1}{n_s!} \sum_{(m_1,..m_{n_s})_n} 
\prod_{j=1}^{n_s}  \int_{-
 \infty}^{+\infty} \frac{dk_j}{2 \pi} 
\prod_{1\leq i < j  \leq n_s} \frac{4(k_i-k_j)^2 +  (m_i - m_j)^2}{4(k_i-k_j)^2 + (m_i + m_j)^2} \\
&&
\prod_{j=1}^{n_s} \frac{1 }{m_j } 
\left( \frac{  \Gamma(-\frac{m_j}{2} + \frac{ \gamma}{2} - i k_j ) }{  \Gamma(\frac{m_j}{2} + \frac{ \gamma}{2} - i k_j )  } \right)^{\frac{t}{2} +1 +x} \left( \frac{   \Gamma(-\frac{m_j}{2} + \frac{ \gamma}{2} +i k_j )}{ \Gamma(\frac{m_j}{2} + \frac{ \gamma}{2} + i k_j ) } \right)^{\frac{t}{2} +1-x} \nn
\eea 
where $\bar{c}$ does not appear explicitly (it appears only via $\gamma$). The dependence of this expression on the variables $(x,t)$ suggests to reintroduce the original coordinates of the square lattice   $ I = \frac{t}{2} +1 +x $ and $J  = \frac{t}{2} +1 -x$ (see Section \ref{Model} and Figure \ref{coord}) and in the following we note $Z(I,J) = Z_{I+J-2}(\frac{I-J}{2} )$.

\bigskip
This formula should be valid for arbitrary $I,J$, and in particular when evaluated for $(I,J)=(1,1)$, for which it should simplify to $\overline{w^n}= \frac{ \Gamma( \gamma - n) }{  \Gamma( \gamma ) }$. Verifying that property is a quite non trivial
check of the procedure (e.g. of the completeness). Although we did not attempt to provide a general proof, we have successfully checked it for various $n$ using Mathematica or the residues theorem (see \ref{appcheck}).

\bigskip
We stress here that this formula is ambiguity-free when the moment problem is well-defined: $ m \leq n \leq \gamma $ and should reproduce all existing moments. 
Very much like what happens for $\overline{w^n}$, it also suggests an analytic continuation, which we use below to derive results on the full probability distribution.

\section{Generating function}\label{GF}

 Our goal is to calculate the Laplace transform of the probability distributions of the partition sum:
\be
g_{I,J}(u) = \overline{ \exp{ - u Z(I,J) } }
\ee
However, as it can be seen already for the one-site problem $I=J=1$,
this Laplace transform must contain two pieces: (i) one that comes from
the {\it generating function of the integer moments} and (ii) a second
piece, which we will conjecture below from an analytic continuation. The one-site problem and the length $2$ polymer are very instructive in that respect and are studied in \ref{app:laplace}.

 \subsection{Generating function for the moments}

Since we only know the integer moments of the partition sum, we start by computing the contribution in $g_{I,J}(u)$ that comes from the moments, i.e. we define the series:

\begin{equation}
  g_{I,J}^{mom}(u) = 1 + \sum_{n =1 }^{ + \infty}(-1)^n \frac{ u^n }{ n!} \overline{Z(I,J)^n}
\end{equation}
where $\overline{Z(I,J)^n}$ denotes in this expression the right hand side of (\ref{momentn3}) for arbitrary integers $n \geq 1$. 
While this distinction is immaterial for $ n < \gamma$, it already implies an analytic continuation since $\overline{Z(I,J)^n}$ does not exist for $n>\gamma$, while the r.h.s. of (\ref{momentn3}) does.

We can use the same strategy as in \cite{we}, \cite{we-flat}.
Since we sum over $n$, the summations over the $n_s$ and the $m_j$ hidden in the expression (\ref{momentn3}) for $\overline{Z(I,J)^n}$ become free summations from $1$ to $\infty$. Permuting the summations over $n$ and over the $m_j$ leads to

\begin{equation}{\label{gener}}
 g_{I,J}^{mom}(u) =  1 + \sum_{n_s =1 }^{ + \infty} \frac{ 1 }{ n_s!} Z(n_s,u)
\end{equation}

with

\bea
\!\!\!\!\!\!\!\!\!\!\!\!\!\!\!\! Z(n_s,u) & = &  \prod_{j=1}^{n_s} \sum_{m_j=1}^{+\infty} \int_{-
 \infty}^{+\infty}  \frac{dk_j}{2 \pi}
\prod_{1\leq i < j  \leq n_s} \frac{4(k_i-k_j)^2 +  (m_i - m_j)^2}{4(k_i-k_j)^2 + (m_i + m_j)^2}   \nonumber \\
&&\prod_{j=1}^{n_s} (-1)^{ m_j}  u^{m_j} \frac{1}{m_j} \prod_{j=1}^{n_s} 
\left( \frac{  \Gamma(-\frac{m_j}{2} + \frac{ \gamma}{2} - i k_j ) }{  \Gamma(\frac{m_j}{2} + \frac{ \gamma}{2} - i k_j )  } \right)^{I} \left( \frac{   \Gamma(-\frac{m_j}{2} + \frac{ \gamma}{2} +i k_j )}{ \Gamma(\frac{m_j}{2} + \frac{ \gamma}{2} + i k_j ) } \right)^{J} \label{Znsu} 
\eea 
and the sums over the $m_j$ are free. 

It is shown in \ref{Fredholmdet} that this expression 
has the structure of a determinant, which allows us to express the generating function as a Fredholm determinant:

\begin{equation}
 g_{I,J}^{mom}(u) = {\rm Det} \left( I  + K_{I,J}^{mom} \right)
\end{equation}
with the kernel:
\begin{eqnarray}\label{firstfredholm}
&&  K_{I,J}^{mom}(v_1,v_2) =  \\
  && \!\!\!\!\!\!\!\!\!\!\!\!\!\!\!\!\!  \sum_{m=1}^{\infty} \int_{-
 \infty}^{+\infty}   \frac{dk}{ \pi}  (-u)^m   e^{ -  2 i k(v_1-v_2) -  m (v_1+v_2) }  \left( \frac{  \Gamma(-\frac{m}{2} + \frac{ \gamma}{2} - i k) }{  \Gamma(\frac{m}{2} + \frac{ \gamma}{2} - i k )  } \right)^{I} \left( \frac{   \Gamma(-\frac{m}{2} + \frac{ \gamma}{2} +i k )}{ \Gamma(\frac{m}{2} + \frac{ \gamma}{2} + i k ) } \right)^{J} \nonumber
\end{eqnarray}
and $ K_{I,J}^{mom} : L^2 ( \mathbb{R}_+) \to L^2 ( \mathbb{R}_+) $, 
so that the two auxiliary integration variables $v_1$ and $v_2$ are positive.
The sum over $m$ is convergent and the result can be expressed in terms of high order hypergeometric functions $_1F_{4t}$ that are meromorphic and well-defined on (almost) all the complex plane, see \ref{apphypergeom}. One can also verify that, at fixed $m$, the integral on $k$ also converges: rewriting the Gamma function using the Pochhammer's symbol leads to simple rational fractions.

The main property of this function $g^{mom}_{I,J}(u)$ is that its coefficient $(-u)^n$ in its Taylor expansion in $u$ 
reproduces $\overline{Z(I,J)^n}/n!$. In \ref{appcheck} we verify this property for small values of $(I,J)$, which is a non trivial test
of the completeness of the Bethe-Brunet eigenstates.

\subsection{Generating function: Laplace transform}

By analogy with the simpler cases studied in \ref{appanalytic} and \ref{app:laplace}, we now conjecture that the full generating function, i.e. the Laplace transform of $P(Z)$ for the
log-gamma polymer, can be computed using a trick inspired by the Mellin-Barnes identity,
leading to our main result:

\begin{equation}
g_{I,J}(u) = \overline{ \exp{ - u Z(I,J) } }= {\rm Det} \left( I + K_{I,J} \right)
\end{equation}

\begin{eqnarray}\label{Fredholmdet2}
 K_{I,J}(v_1,v_2) = && \int_{-\infty}^{+\infty}   \frac{dk}{ \pi}  \frac{-1}{2i} \int_C \frac{ds}{ \sin( \pi s ) }   u^s  e^{ -  2 i k(v_1-v_2) -  s (v_1+v_2) } \\
 &&  \left( \frac{  \Gamma(-\frac{s}{2} + \frac{ \gamma}{2} - i k ) }{  \Gamma(\frac{s}{2} + \frac{ \gamma}{2} - i k )  } \right)^{I} \left( \frac{   \Gamma(-\frac{s}{2} + \frac{ \gamma}{2} +i k )}{ \Gamma(\frac{s}{2} + \frac{ \gamma}{2} + i k ) } \right)^{J} \nonumber 
\end{eqnarray}
where $C = a + i \mathbb{R}$ with $0<a<1$ (here the sum runs from $1$ to infinity) and $K_{I,J} : L^2 ( \mathbb{R}_+) \to L^2 ( \mathbb{R}_+) $\footnote{ Note that for $I=J$ and $s, ik$ on the imaginary axis the ratio of gamma function is a complex number of modulus unity. For
$a>0$ is has modulus smaller than one, decaying to zero for large $|s|,k$. 
The exponential convergence in $s$ is ensured by the $1/sin$ but the convergence in $k$ 
is slower (algebraic).}. Note that the symmetry $I \leftrightarrow J$ is explicit
under the change of variable $k \leftrightarrow -k$.
We discuss below in Section \ref{Comparison} the connection between this result, obtained via the Bethe Ansatz, and the
previous formula of \cite{logboro}, obtained using a completely different route. 

\subsection{Probability distribution}

Before turning to the large-length limit, let us briefly mention that one can directly obtain from~(\ref{Fredholmdet2}) the probability distribution of $\log Z(I,J)$ as a convolution: $\log Z(I,J) = \log Z_0 + \log \tilde Z(I,J)$ where $- \log Z_0$ is an independent random variable with a standard (unit) Gumbel distribution and $\tilde Z(I,J)$ is distributed according to a probability density $\tilde{P}_{IJ}$ given by 
\bea
\tilde{P}_{I,J}(v) = \frac{1}{2 i \pi v} \left(  {\rm Det} ( I + \check{K}_{I,J}^{(1)} - i  \check{K}_{I,J}^{(2)}  ) - {\rm Det} ( I +  \check{K}_{I,J}^{(1)}  + i  \check{K}_{I,J}^{(2)}  )  \right)
\eea
where $\check{K}_{I,J}^{(j)} $, $j=1,2$, are two operators $\check{K}^{(j)}_{I,J} : L^2 ( \mathbb{R}_+) \to L^2 ( \mathbb{R}_+) $ with kernels:
\begin{eqnarray}\label{probaanytime}
\check{K}_{I,J}^{(j)}(v_1,v_2) = && \int_{-\infty}^{+\infty}   \frac{dk}{ \pi}  \frac{-1}{2i} \int_C \frac{ds}{ f^{(j)}( \pi s ) }   v^{-s} e^{ -  2 i k(v_1-v_2) -  s (v_1+v_2) } \\
 &&  \left( \frac{  \Gamma(-\frac{s}{2} + \frac{ \gamma}{2} - i k ) }{  \Gamma(\frac{s}{2} + \frac{ \gamma}{2} - i k )  } \right)^{I} \left( \frac{   \Gamma(-\frac{s}{2} + \frac{ \gamma}{2} +i k )}{ \Gamma(\frac{s}{2} + \frac{ \gamma}{2} + i k ) } \right)^{J} \nonumber 
\end{eqnarray}
where $f^{(1)}(x)= \tan x$ and $f^{(2)}(x)= 1$. The derivation of this result is given in \ref{app:proba}.

\section{Limit of very long polymers and universality}\label{Asymptotic}

In this section we show how the above formula leads to Tracy-Widom universality and derive explicit expressions
for the asymptotic probability distribution of the free energy. 

Let us consider the large length limit, for which we find more convenient to use our coordinates $(x,t)$ (see Fig.~\ref{coord}), and focus first on the scaling $x \sim \varphi t$ with $-\frac{1}{2}<\varphi<\frac{1}{2}$. We define the free energy as:
\bea
F_t(\varphi) = - \ln Z_{t}(x=\varphi t)
\eea
We thus need to analyze the $t \to \infty$ limit of $g_{\varphi,t}(u) = {\rm Det} \left( I  + K_{\varphi,t} \right)$ with $K_{\varphi,t} : L^2 ( R_+) \to L^2 ( R_+) $ defined by its kernel (from (\ref{Fredholmdet2})):

\begin{eqnarray}\label{Fredholm3asympt}
\!\!\!\!\!\!\!\!\!\!\!\!\!\!\!\!\!\!\!\!\!\!\!\!\!\!\!\!\! K_{\varphi,t}(v_1,v_2) = && \int_{\mathbb{R}}   \frac{dk}{ \pi}  \frac{-1}{2i} \int_C \frac{ds}{ \sin( \pi s ) }   u^s  e^{ -  2 i k(v_1-v_2) -  s (v_1+v_2) }   \\
 && \left( \frac{ \Gamma( -\frac{s}{2} + \frac{\gamma}{2} - ik ) }{ \Gamma( \frac{s}{2} + \frac{\gamma}{2} - ik )  } \right) ^{ 1+t(\frac{1}{2} + \varphi)} \left( \frac{ \Gamma(- \frac{s}{2} + \frac{\gamma}{2} + ik ) }{ \Gamma( \frac{s}{2} + \frac{\gamma}{2} + ik ) } \right) ^{1+t(\frac{1}{2} - \varphi)} \nonumber 
\end{eqnarray}
The behavior of the large length limit is estimated through a saddle-point analysis. We define $G_{\varphi}(x) = (\frac{1}{2} + \varphi) \log \Gamma(\frac{\gamma}{2} -x )-(\frac{1}{2} - \varphi) \log \Gamma(\frac{\gamma}{2} +x )$ to write the Gamma function factor as
\begin{equation} \label{Gphi} 
 \exp{\left( t \left( G_{\varphi}(\frac{s}{2} + ik )-G_{\varphi}(-\frac{s}{2} +ik ) \right)+2 \left( G_{0}(\frac{s}{2} + ik )-G_{0}(-\frac{s}{2} +ik ) \right) \right)}
\end{equation}

  We now use a Taylor expansion around the critical-point $(s , k) = (0 , -i  k_\varphi)$\footnote{this is natural since $\varphi\neq 0$ breaks the symmetry $k \to -k$ of (\ref{Gphi}) while the factor in the exponential remains odd in $s$ .} :
\begin{equation}\label{dvlpt}
 G_{\varphi}(\frac{s}{2} + ik )-G_{\varphi}(-\frac{s}{2} +ik ) = 0 +  G_{\varphi}'( k_\varphi ) s  + G_{\varphi}''( k_\varphi ) i s \tilde k + \frac{ G_{\varphi}'''( k_\varphi )}{6} (\frac{s^3}{4} -3 s \tilde k^2) + O(s^4)
\end{equation}
where $\tilde k = k + i k_\varphi$ and $s$ are considered to be of the same order (this is indeed the case, see below). It is easy to see that $G_{\varphi}'( k_\varphi )$ corresponds to the additive part of the free-energy. This is thus the proper saddle-point only if $G_{\varphi}''( k_\varphi )$ is $0$, which implicitly defines $k_\varphi$ as a function of $\varphi$ as the solution of the equation:
\begin{equation}\label{saddlepoint}
 (\frac{1}{2} + \varphi) \psi'(\frac{\gamma}{2} - k_\varphi)-(\frac{1}{2} - \varphi) \psi' (\frac{\gamma}{2} +k_\varphi )=0
\end{equation}
where $\psi= \frac{ \Gamma'}{\Gamma}$ is the digamma function. The numerical solution $k_\varphi$ is plotted in \ref{appnumerics}. The expansion (\ref{dvlpt}) indicates that we have to rescale the free-energy 
 as:
\begin{equation}  
 F_t(\varphi) =  c_{\varphi} t + \lambda_{\varphi} f_t
\end{equation}
where $c_{\varphi}= -G_\varphi'(k_\varphi)$ is the free-energy per unit length (which is self-averaging at large $t$)
and $\lambda_{\varphi} = \left( \frac{ t G_{\varphi}'''( k_\varphi ) }{8} \right)^{\frac{1}{3}}$ is the scale of the
free energy fluctuations, such that $f_t$ is an $O(1)$ random variable. 
With these definitions, the rescaled generating function of the $\lambda_{\varphi}$ rescaled free energy, $\tilde g_{\varphi,t}(z ) = 
\overline{ \exp( - e^{ -\lambda_{\varphi}( z+ f_t) } ) }$, is given by the Fredholm determinant of a rescaled kernel, $\tilde g_{\varphi,t}(z )  = {\rm Det}(I+ \tilde{K}_{\varphi,t} )$, which is obtained by rescaling $s \to \frac{s}{\lambda_{\varphi}} $, $ \tilde k  \to \frac{\tilde k}{\lambda_{\varphi}} $, as well as $v_i \to \lambda_{\varphi} v_i$:
\begin{equation}
\tilde{K}_{\varphi , t}(v_1,v_2) = \int_{\mathbb{R}} \frac{d \tilde k}{ \pi} \frac{-1}{2i} \int_C \frac{ds}{\lambda_{\varphi} \sin( \pi \frac{s}{\lambda_{\varphi}} ) }   e^{  -s z -  2 i \tilde k  (v_1-v_2) -  s(v_1+v_2)  - 4 \tilde k^2 s + \frac{s^3}{3}+O(\frac{1}{\lambda_{\varphi}})  }  
\end{equation}
 where the $O(\frac{1}{\lambda_{\varphi}})$ term contains higher order derivatives of $G_{\varphi}$ and the expansion of $G_0$ around $k_{\varphi}$~\footnote{ The extra factor $e^{-2 k_\varphi \lambda_\varphi (v_1-v_2)}$ originating from the change of variable has been
removed since it is immaterial in the calculation of the Fredholm determinant.}. The large polymer length limit $\lambda_{\varphi} \to \infty$ can be safely taken in this last expression, leading to a kernel $\tilde{K}_{\infty}$ for which there is more freedom in the choice of the integration contour $C$: it should only define a convergent integral and passes to the right of zero. The $t\to\infty$ limit of the rescaled generating function can thus be written as $\lim_{t \to \infty} \tilde g_{\varphi,t}(z )  =  Prob(-f <z) = {\rm Det}(I +\tilde{K}_{\infty} ) $ with
\begin{equation}
\tilde{K}_{\infty}(v_1,v_2) = \int_{\mathbb{R}}   \frac{d \tilde k}{ \pi} \int_C \frac{-ds}{2 i \pi  s }  e^{  -s z -  2 i \tilde k  (v_1-v_2) -  s(v_1+v_2)  - 4 \tilde k^2 s + \frac{s^3}{3}  }  
\end{equation}
which corresponds to the Tracy-Widom GUE distribution. Indeed, the Airy trick $\int_{\mathbb{R}} dy Ai(y) e^{ys} = e^{\frac{s^3}{3}}$ valid for $Re(s)>0$, followed
by the shift $y \to y + z + v_1 + v_2 + 4 \tilde k^2$, the identity $\int_C \frac{ds}{2 i \pi s} e^{s y} =  \theta(y) $, and the rescaling $\tilde k \to \tilde k/2$ give
\begin{equation}
\tilde{K}_{\infty} (v_1,v_2) =  - \int_{\mathbb{R}}   \frac{d\tilde k}{ 2 \pi} \int_{\mathbb{R}_{+}} dy   Ai(y  + z + v_1 + v_2 +  \tilde k^2) e^{ - i \tilde k  (v_1-v_2)  } 
\end{equation}
which is one way to define $F_2$ as in \cite{we} : this kernel indeed corresponds to $Prob(-f_{\infty} <z) =\det( I  + \tilde K_{\infty})  = F_2(2^{-\frac{2}{3}} z) $. Putting everything together, our result for the asymptotic limit reads
\begin{equation}\label{asymptoticlim}
\lim_{t \to \infty} Prob\left( \frac{ \log Z_t( \varphi t) + tc_{\varphi}}{\lambda_{\varphi} } <2^{\frac{2}{3}} z \right) = F_2(z)
\end{equation}
where $F_2(z)$ is the standard GUE Tracy-Widom cumulative distribution function, and the (angle-dependent)
constants are determined by the system of equations:
\begin{eqnarray}
&&0=(\frac{1}{2} + \varphi) \psi'(\frac{\gamma}{2} - k_\varphi)-(\frac{1}{2} - \varphi) \psi' (\frac{\gamma}{2} +k_\varphi )\\
&&c_{\varphi}= (\frac{1}{2} + \varphi) \psi(\frac{\gamma}{2} - k_\varphi)+(\frac{1}{2} - \varphi) \psi (\frac{\gamma}{2} +k_\varphi )\\
&&\lambda_{\varphi}=\left( -\frac{t}{8} \left( (\frac{1}{2} + \varphi) \psi''(\frac{\gamma}{2} - k_\varphi)+(\frac{1}{2} - \varphi) \psi''(\frac{\gamma}{2} +k_\varphi ) \right) \right)^{\frac{1}{3}}
\end{eqnarray}
\paragraph{Central region (i.e. square lattice diagonal):}  In the special case $\varphi=0$ the solution is explicit: $k_{\varphi} =0$  and the free energy per unit length and the scale of the free-energy fluctuations are given by
\begin{equation}
\lambda_{0}  = (- t \frac{ \psi''( \frac{\gamma}{2} ) }{8} )^{\frac{1}{3}}  \quad  c_0= \psi(\frac{\gamma}{2})
\end{equation}
For small angle $\varphi$ one can also compute pertubatively the first correction, which is $k_{\varphi} = \frac{2 \psi'(\frac{\gamma}{2})}{ \psi''(\frac{\gamma}{2})} \varphi +O(\varphi^3)$. This allows to obtain the leading correction to the extensive part of the mean-free energy as a function of the
angle, and of the endpoint position, as:
\be
t c_{\varphi} = t \psi( \frac{\gamma}{2}) -t \frac{ 2 \psi'(\frac{\gamma}{2}) ^2}{\psi''(\frac{\gamma}{2})} \varphi^2 + O(\varphi^4)
=  t \psi( \frac{\gamma}{2}) - \kappa \frac{x^2}{4 t} + .. 
\ee
which defines the effective elastic constant $\kappa$ as (the last equation is valid in the scaling region $x/t \ll 1$)
\be
\kappa = \left( -8 \frac{ \psi'(\frac{\gamma}{2}) ^2}{ \psi''(\frac{\gamma}{2}) }\right)
\ee
We see here that, although the discrete model does not obey an exact statistical tilt symmetry (STS), see e.g. \cite{we-flat},
this symmetry is recovered at large scale (within this scaling region) with an effective elastic constant originating
from the geometrical entropy effect. 

\paragraph{Remark on the digamma function} The appearance of the digamma function in the mean free energy is natural since, as was noted in \cite{logsep1}, a potential $V= - \ln w$ distributed according to a log-Gamma distribution of parameter $\gamma$ verify $\overline{V^q} = \partial_\gamma^{q-1} \psi(\gamma)$. However, the appearance of the parameter $\frac{\gamma}{2}$ is non-trivial and has to do with the existence of an invariant measure of parameter $\frac{\gamma}{2}$ as was proved in \cite{logsep1} using peculiar boundary conditions. Here we did not use these boundary conditions and this is visible in the fact that $\lim_{\varphi \to \frac{1}{2} } c_\varphi = \psi(\gamma)$
(see \ref{appnumerics}): when one approaches the border of the lattice one retrieves the original parameter $\gamma$ since there is a single path. The behavior of the above equations is, however, ill-defined in this limit: this is a signature that, at $\varphi= \frac{1}{2}$, the fluctuations of the free-energy become Gaussian and scale as $\sqrt{t}$ (as a simple application of the central limit theorem).

\paragraph{Lieb-Liniger limit}

We can recover the results of \cite{we,dotsenko,spohnKPZEdge,corwinDP} in the continuum (Lieb-Liniger) limit by considering 
the LL limit (see Section \ref{LLlimit}) around the angle zero (since in that limit $x/t \sim {\sf a}$). 
Using $\psi( x ) \sim_{x \to \infty} \log x -\frac{1}{2x}-\frac{1}{12 x^2} + O(x^{-4}) $ and (\ref{LLcorresponence}) one can show the following Lieb-Liniger limits:
\begin{eqnarray}
&& \lambda_0 =  (- t \frac{ \psi''( \frac{\gamma}{2} ) }{8} )^{\frac{1}{3}} \simeq_{LL} (\frac{{\bar{c}_{LL}}^2t_{LL}}{4})^{\frac{1}{3}}\\
&& c_0 t \simeq_{LL}  \frac{8}{\sf a^2} t_{LL} \ln(\frac{2}{{\sf a} \bar c_{LL}}) + \frac{\bar c_{LL} }{12} t_{LL} + O({\sf a}^2) \quad , \quad  \kappa \frac{x^2}{4 t} \simeq_{LL} \frac{x_{LL}^2}{4 t_{LL}} \nonumber
\eea
where the first term in the extensive part of the mean free energy arises from lattice entropic effect and
can be anticipated from (\ref{LLcorresponence}). Putting all together, one recovers the
result for the one point distribution of the continuum Airy$_2$ process:
\begin{equation}\label{asymptoticlim2}
\lim_{t_{LL} \to \infty} Prob\left( \frac{ \log Z^{LL}_{t_{LL}}(x_{LL}) + \frac{x_{LL}^2}{4 t_{LL}} + \frac{ \bar c_{LL}}{12} t_{LL}}{(\frac{{\bar{c}_{LL}}^2t_{LL}}{4})^{\frac{1}{3}} } <2^{\frac{2}{3}} z \right) = F_2(z)
\end{equation}

\paragraph{Numerical results:} Using a direct simulation of (\ref{evolutionequation}) with Mathematica, we
calculate the partition sum for various lengths and samples of environments. This provides some numerical verifications of the above results.
The full check of (\ref{asymptoticlim}) is qualitatively satisfying. In Fig \ref{figfullconvergence} we show the convergence of the two first cumulants of the probability distribution of $F_t(0)$ for $\gamma=3$ and $t=2^i$, $i=1,...,13$. Numerical cumulants are evaluated using $N=10^5$ samples ($i=1,...,10$) or $N=10^4$ ($i=11,12,13)$. The mean free energy $\frac{\overline{F_t(0)}}{t}$ quickly converges since the theoretical prediction (\ref{asymptoticlim}) already includes a finite size correction. The asymptote is $\psi(\gamma/2)=0.03649$.
The convergence of the rescaled variance $\frac{Var(F_t(0))}{2^{\frac{4}{3}}\lambda_0^2}$ is slower but in good agreement with the Tracy-Widom asymptotic value $0.813$. 
\begin{figure}[H]
   \centering
   \includegraphics[scale=0.4]{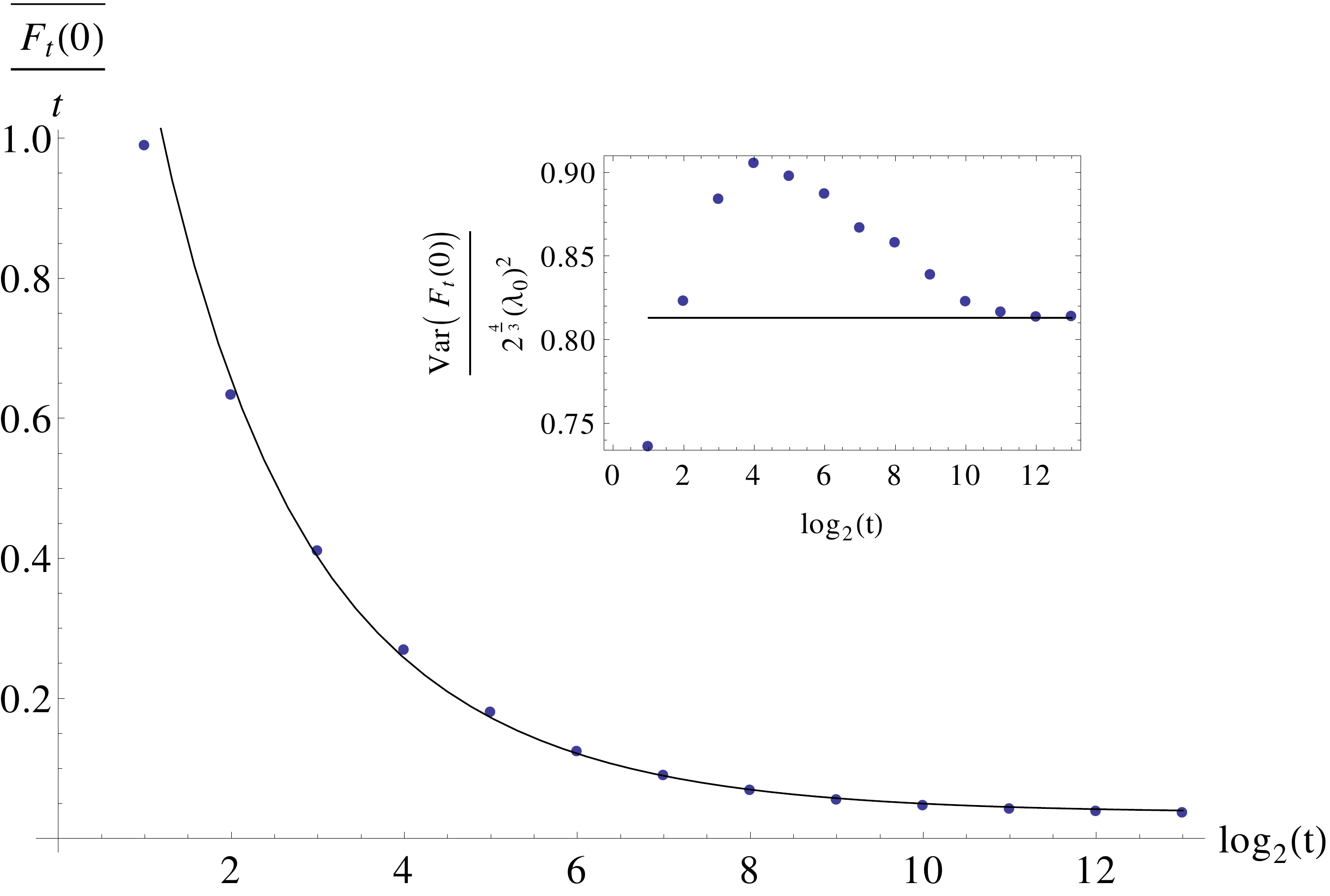}
   \caption{Convergence of the mean free energy density $\frac{\overline{F_t(0) }}{t} $ (main curve) and of the rescaled variance of the free-energy $ \frac{Var(F_t(0) )}{2^{\frac{4}{3}} (\lambda_0)^2}$ (inset) as compared to the theoretical prediction (\ref{asymptoticlim}). The blue dots are the numerical results, the black lines are the theoretical predictions of (\ref{asymptoticlim}). There are no fitting parameter.}
   \label{figfullconvergence}
\end{figure}

We also checked the dependence on $\varphi$ of the two first rescaled cumulants. In Fig \ref{figfullspatialdependance} we show the obtained dependence of $\frac{\overline{F_t(\varphi)}}{t}$ and $\frac{Var(F_t(\varphi))}{t^{\frac{2}{3}}}$ for $\gamma=3$ and $t=4096$. These cumulants are numerically evaluated using $10^4$ samples. The theoretical predictions are given by (\ref{asymptoticlim}) where $k_{\varphi}$ is evaluated as explained in \ref{appnumerics}.
\begin{figure}[H]
   \centering
   \includegraphics[scale=0.4,trim=0 0 0 0, clip]{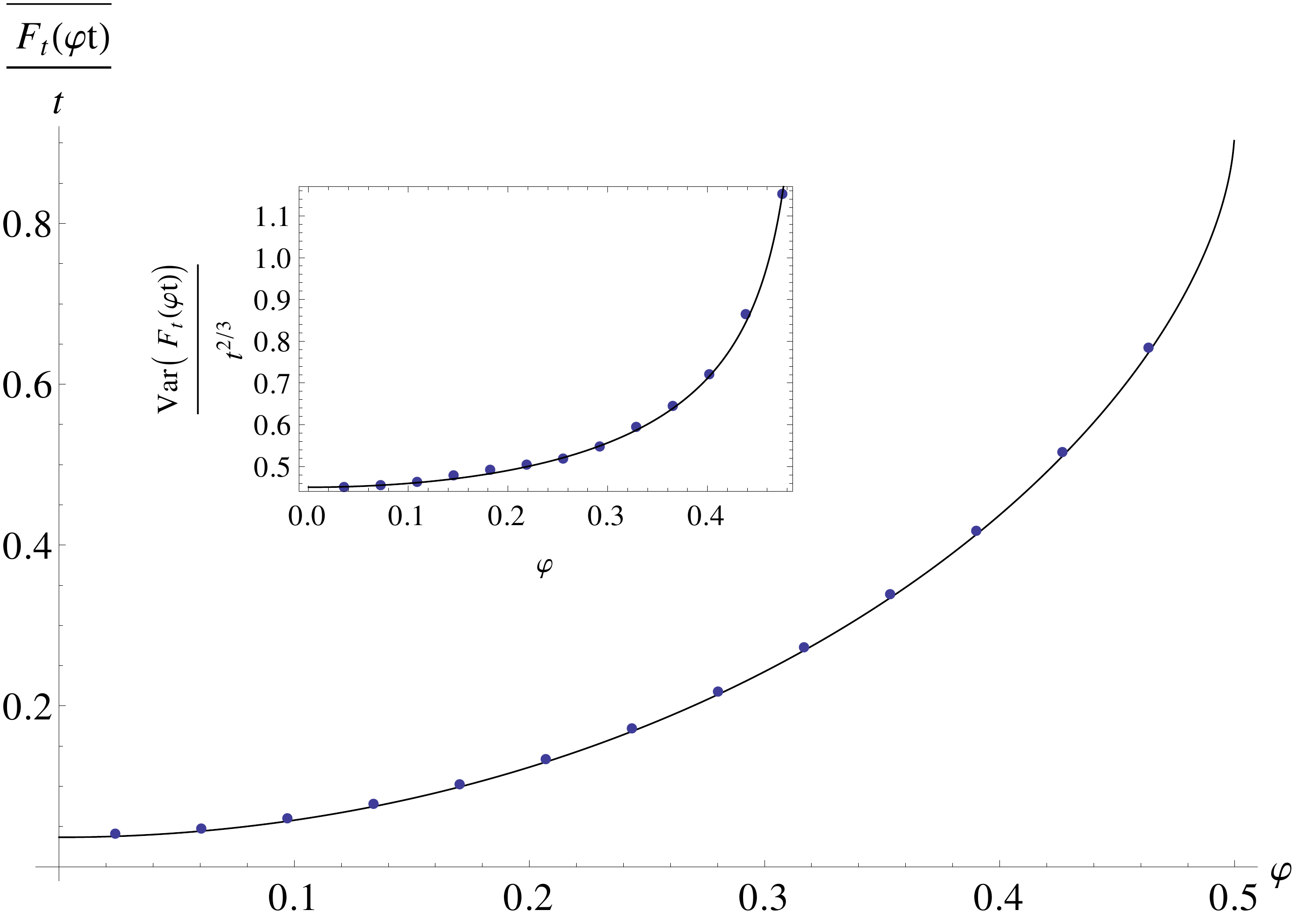}
\caption{Dependence on the endpoint position $x= \varphi t$ of the mean free energy density $\frac{\overline{F_t(\varphi t) }}{t} $ (main curve) and of the rescaled variance of the free-energy $ \frac{Var(F_t(\varphi t) )}{t^{\frac{2}{3}}}$ (inset), for $t=4096$. The blue dots are the numerical results, the black lines are the theoretical predictions from (\ref{asymptoticlim}). There are no fitting parameter.}
   \label{figfullspatialdependance}
\end{figure}

\paragraph{Semi-discrete O'Connell-Yor polymer}
Let us finally mention another interesting asymptotic limit that is briefly discussed in \ref{appsemi} and that allows us to retrieve the semi-discrete directed polymer model of \cite{semidiscret1}. This limit is most conveniently studied on the equivalent form (\ref{finalkernel}) of the Fredholm determinant formula (\ref{Fredholmdet2}) that is derived in the next section.

\section{Comparison with other results}\label{Comparison} 

\paragraph{Mathematical Results}

Using the geometric RSK correspondence, it was shown in \cite{logsep2}, that the Laplace transform of the partition sum of the polymer with fixed endpoints $(1,1) \to (I,J)$ with $I \geq J$ can be expressed as a J-fold integral:

\bea
\!\!\!\!\!\!\!\!\!\!\!\! \overline{e^{- u Z(I,J) }} = \frac{1}{J !} \int_{(i R)^{J}} \prod_{j=1}^{J} \frac{dw_j}{2 i \pi} \prod_{j \neq k =1}^{J} \frac{1}{\Gamma(w_j-w_k)} 
[ \prod_{j=1}^{J} u^{w_j-a}  \Gamma[a-w_j]^{J} \frac{\Gamma(\alpha-w_j)^{I}}{\Gamma(\gamma)^{I}} ]
\label{sepp1} 
\eea 
where $\alpha-a = \gamma >0 $, the parameter of the underlying inverse Gamma distribution. In \cite{logboro}, it was shown that this integral can be expressed as a Fredholm determinant: $\overline{e^{- u Z(I,J)}} = {\rm Det}( I + K_{I,J}^{RSK} ) $ with
\bea \label{borodinK} 
 \!\!\!\!\!\!\!\!\!\!\!\! K_{I,J}^{RSK}(v,v') =  \frac{1}{ (2 \pi i)^2} \int_{ l_{\delta_2}} dw \frac{\pi}{\sin( \pi(v - w) ) } \frac{1}{ w -v'} u^{w-v} \left( \frac{ \Gamma( \alpha- w) }{ \Gamma( \alpha - v)} \right) ^{I} \left( \frac{\Gamma( v - a)}{\Gamma( w - a) } \right) ^{J}
\eea
where $0<\delta_2 <1$ , $ 0 < \delta_1 < min \{ \delta_2 , 1- \delta_2 \}$ and $0 < a<\delta_1$, $\alpha>\delta_2$. Here $l_{\delta_2}$ denotes the axis  $ Re(z) = \delta_2 $ oriented from the bottom to the top. $K$ is the kernel of an operator $ L^2( C_{\delta_1} ) \rightarrow L^2( C_{\delta_1} ) $ with $C_{\delta_1}$ a positively oriented circle of center $0$ and radius $ \delta_1 $. The measure of integration on $C_{\delta_1}$ is chosen here as the Lebesgue measure, hence the
extra factor of $1/(2 i \pi)$ as compared to \cite{logboro} that uses a different convention. The contour for the $v,v'$ integrals is tailored so that only the pole at $v=a$ contributes.
Using this expression, they could perform the asymptotic analysis and show that
\bea \label{Asymptoticproba}
\lim_{N \to \infty}  Proba \left( \frac{ log( Z(N,N)) +2 N \psi( \frac{\gamma}{2})  }{(N)^{ \frac{1}{3}}}  <  (- \psi''( \frac{\gamma}{2}) ) ^{ \frac{1}{3}}  z \right) = F_2 ( z)
\eea
which is exactly the same result as ours in (\ref{asymptoticlim}) for the case of the central region $\varphi=0$. 

\paragraph{Kernels correspondence} 
We now sketch how we find the kernel $K_{I,J}^{RSK}$ and our kernel to be closely related. We start from our result (\ref{Fredholmdet2})
where $C = a + i \mathbb{R}$ with $0<a<1$. The first step is to make the change of variables $s = a + i \tilde{s}$, which allows us to rewrite this kernel as an integral on $\mathbb{R}^2$:
\begin{eqnarray}
 K_{I,J}(v_1,v_2) = && \int_{\mathbb{R}^2}   \frac{-dk d\tilde{s}}{ 2 \pi}  \frac{1}{ \sin( \pi ( a + i \tilde{s}) ) }   u^{ a + i \tilde{s}}  e^{ -  2 i k(v_1-v_2) -  ( a + i \tilde{s}) (v_1+v_2) } \\
 &&   \left( \frac{ \Gamma( -\frac{ a + i \tilde{s}}{2} + \frac{\gamma}{2} - ik )  }{ \Gamma( \frac{ a + i \tilde{s}}{2} + \frac{\gamma}{2} - ik )  } \right) ^{I}  \left( \frac{ \Gamma( -\frac{ a + i \tilde{s}}{2} + \frac{\gamma}{2} + ik ) }{  \Gamma( \frac{ a + i \tilde{s}}{2} + \frac{\gamma}{2} + ik ) } \right) ^{ J} \nonumber
\end{eqnarray}
We now use the change of variables $(k , \tilde{s} ) \to (s_+ ,s_-) $ with $s_+ =\frac{ \tilde{s}}{2} + k$ and  $s_- = \frac{\tilde{s}}{2} - k$, this gives
\begin{eqnarray}
&& K_{I,J}(v_1,v_2)  =   \int_{R^2} ds_+ ds_- A( v_1 , s_+) B(s_+ , s_-) C(s_- , v_2)
 \end{eqnarray}
where we introduced  $ \gamma_- = \gamma -a   $  and $ \gamma_+ = \gamma +a$ and
  \begin{eqnarray}
 \!\!\!\!\!\!\!\!\!\!\!\!\!\!\!\!\! A( v_1 , s_+)   =   e^{-v_1 (2 i s_+ +a ) } \quad  \quad C(s_- , v_2) = e^{-v_2 (2 i s_- +a ) } \\
\!\!\!\!\!\!\!\!\!\!\!\!\!\!\!\!\! B(s_+ , s_-)  = \frac{-1}{2 \pi} \frac{1}{ \sin( \pi ( a + i(s_+ +s_-)) ) }   u^{ a + i (s_+ + s_-) }  \left( \frac{ \Gamma( \frac{\gamma_-}{2}  - i s_+ ) }{  \Gamma(\frac{\gamma_+}{2} + i s_-) } \right) ^{ I} \left( \frac{\Gamma( \frac{\gamma_-}{2} - i s_- ) }{ \Gamma( \frac{\gamma_+}{2} +  i s_+) } \right) ^{ J } 
 \end{eqnarray}
The kernel now has the form of a product of operators, hence we can use the identity $ {\rm Det}( I + ABC ) = {\rm Det}( I + CAB) $ (from the cyclic property of the trace) to obtain that the Laplace transform $g_{I,J}(u)$ can be expressed as the Fredholm determinant $g_{I,J}(u)={\rm Det}( I + K''_{I,J} )$ with $K''_{I,J}  = CAB $:
\begin{eqnarray}
\!\!\!\!\!\!\!\!\!\!\!\!\!\!\!\!\!  K''_{I,J}( v , v') & = & \int_{R} ds_+ \int_{R_+} dv_2  C( v , v_2) A(v_2 ,s_+) B (s_+, v')
\end{eqnarray}
where in this expression, the integral on $v_2$ is straightforward and we find
\begin{eqnarray}
 K''_{I,J}( v , v') = && \int_{R} ds_+ \frac{-1}{4 \pi (a + i (s_+ + v))} \frac{1}{ \sin( \pi ( a + i(v'+s_+)) ) }   u^{ a + i (v'+  s_+) }  \\
 && \left( \frac{ \Gamma( \frac{\gamma_-}{2}  - i s_+ ) }{  \Gamma(\frac{\gamma_+}{2} + i v') } \right) ^{ I} \left( \frac{\Gamma( \frac{\gamma_-}{2} - i v' ) }{ \Gamma( \frac{\gamma_+}{2} +  i s_+) } \right) ^{ J } \nonumber  
\end{eqnarray}
where now $K''_{I,J} : L^2 (R) \to L^2(R) $. Note that the convergence of the integral over $s_+$, a necessary 
condition to exchange the integrations, is satisfied \cite{foot4} when $I \geq J$, which we now assume. If $J \geq I$ we would instead write $ {\rm Det}( I + ABC ) = {\rm Det}( I + BCA) $, leading to the same kernel with $I$ and $J$ exchanged (and $v,v'$ exchanged which is immaterial in the Fredholm determinant). Using the change of variables $w = a+\frac{\gamma_+}{2} + is_+$ and $ z=-iv' + \frac{\gamma_+}{2}$ (it adds a minus sign),  the result for $g_{I,J}(u)$ is re-expressed as the Fredholm determinant 
$g_{I,J}(u)={\rm Det}( I + K'_{I,J} )$ with $K'_{I,J} : L^2 (\frac{\gamma_+}{2} + iR) \to L^2(\frac{\gamma_+}{2}+ iR) $ and
\begin{eqnarray}\label{Fredholm4}
 \!\!\!\!\!\!\!\!\!\!\!\!\!\!\!\!\!\!\!\!\!\!\!\!\!\!\!\!\!\!\!  K'_{I,J}(z , z') = \int_{a + \frac{\gamma_+}{2}+ i R} d w 
\frac{1}{4 \pi ({w-z'})} \frac{1}{ \sin( \pi ( w- z) ) }   u^{  w- z }   \left( \frac{\Gamma(\gamma+ a - w ) }{ \Gamma(\gamma+ a - z) } \right) ^{ I}     \left( \frac{ \Gamma(  z - a ) }{  \Gamma( w - a) }\right) ^{ J} 
\end{eqnarray}
In this last expression we have some freedom in the choice of the contours: the evaluation of the Fredholm determinant involves integrals on $w$ and on $z$ that are invariant as long as we translate the contours of integration by the same amount, and that we do not cross the poles located at $w=\gamma+a + n$ and $z=a-n$ for $n \in \mathbb{N}$. We can thus write our final result as $g_{I,J}(u) = {\rm Det}(I +K_{I,J}^{BA})$ with $K_{I,J}^{BA} : L^2(a+\tilde{a} + i \mathbb{R}) \to L^2(a+\tilde{a}+ i\mathbb{R})$ defined as the "Bethe ansatz" kernel:
\begin{equation}\label{finalkernel}
K_{I,J}^{BA}(z , z') = \int_{2 a+ \tilde{a} + i \mathbb{R} } d w 
\frac{1}{4 \pi ( w-z')} \frac{1}{ \sin( \pi ( w- z) ) }   u^{  w- z }  \left( \frac{\Gamma(\alpha - w ) }{ \Gamma(\alpha - z) } \right) ^{ I } \left( \frac{ \Gamma(  z - a ) }{  \Gamma( w - a) }\right) ^{ J}    
\end{equation}
where $\alpha= \gamma  +a$, $0<a<1 $ and $0<\tilde a<\gamma-a$ and $I \geq J$.
\paragraph{}
The next step to achieve the correspondence would be to deform the contour of integration of $z$ into the circle $C_{\delta_1}$. This seems to be a difficult task since when deforming the contour one a priori encounters 
an infinite number of poles. However we conjecture that it works and that:
\be
{\rm Det}(I+K_{I,J}^{BA}) = {\rm Det}(I+K_{I,J}^{RSK})
\ee
We verified that identity in some simple cases, e.g. 
by explicitly computing the $u$, $u^2$, $u^3$ terms (for $t=0$ and $t=2$) and the $u^{\gamma}$,  $u^{\gamma+1}$ terms ($t=0$ only). 
A proof may require lifting the model to a higher generalization involving Macdonald processes \cite{Borodinprivate}. Note that in the case of the semi-discrete polymer (see \ref{appsemi}), the equivalence between such small circles and open contours was already proved in \cite{BCF}.

Let us finally mention that this kernel allows to obtain another formula for the probability distribution of $\log Z$ 
analogous to (\ref{probaanytime}). More precisely (\ref{probaanytime}) still holds
with $\check{K}_{I,J}^{(j)}  \to \check{K}_{I,J}^{(j),BA}$ where the kernels $\check{K}_{I,J}^{(j),BA}$
are obtained from $K_{I,J}^{BA}$ in (\ref{finalkernel}) by substituting~$\frac{1}{ \sin( \pi ( w- z) ) }   u^{ w-z } \to v^{z-w}/f^{(j)}( \pi(w-z) )$.
\paragraph{Results from the physics literature }
During the last stage of the redaction of this article, we became aware of a very recent work \cite{povolo} 
where zero-range $q$-boson models with factorized steady state measures and which are integrable via
the Bethe ansatz are classified. Although these results were obtained in a different context, there is a clear connection 
to the ansatz studied here. The main difference is that the stochasticity hypothesis has to be relaxed to get a more general framework 
that encompasses our model. This is however easily done (work in progress) 
and the Brunet ansatz then appears as a (singular) limit of this 
generalized ansatz.

\section{Conclusion}\label{Concl}

In this paper we have studied the problem of a directed polymer on the square lattice 
in presence of log-Gamma distributed quenched random weights. Building up on
an earlier work by Brunet, we have shown how the Bethe-Ansatz and integrability techniques could be 
efficiently used to derive an exact formula for the $n$-th integer moment 
of the partition function for fixed endpoints and arbitrary polymer length, (\ref{momentn3}), defined for $n<\gamma$.
Based on this formula and the observations made in \ref{appanalytic} and \ref{app:laplace}, we conjectured a formula for the Laplace transform of the probability distribution of the partition sum. From this: (i) we obtained a formula for the probability distribution of the partition function for any polymer length (\ref{probaanytime}) (ii) we showed convergence of the free energy distribution 
to the Tracy-Widom distribution at large time (\ref{asymptoticlim}) and derived the normalizing constants
and their dependence in the endpoint position (i.e. in the angle with respect to the diagonal of the lattice). 
Specifically, we obtained the extensive part of the mean free energy, as well as the variance of
the fluctuations. From the angle dependence we also obtained the elastic coefficient.
We performed numerical simulations of long polymers to check and confirm
some of these results with very good agreement. At each stage of the calculation we proved that
all of our formulas reduce, in the continuum limit, to the ones for the Lieb-Liniger model, thereby
recovering the results for the continuum KPZ model obtained in previous works.

In the last section we showed how these results are related to the previous work of \cite{logboro}. 
 Our asymptotic limit agrees and extends their result to arbitrary angle, and our Fredholm determinant formula are closely related, with an essential difference in the contours of integration. This difference seems to be a signature of the method: our integrability techniques naturally give rise to "large" contours formulas, whereas the techniques used in the mathematical context give rise to "small" contours formulas. 
 Although we provided some verifications, the full proof of the equivalence of the two formulas
 may require considering a regularized, (e.g. q-deformed) version of the log-Gamma model
 \cite{Borodinprivate}. 

\paragraph{} 
This paper thus offers new tools which could be used to explore further the similarities between quantum integrability 
and tropical combinatorics methods. It also opens the way to other studies on the log-Gamma directed polymer with e.g. 
other boundary conditions, such as flat (as in \cite{we-flat}) or stationary
(as in \cite{logsep1}) and extension to the inhomogeneous model of \cite{logsep2}, which are left
for future studies. 
%
%
%

\bigskip

{\it Acknowledgments:}

We want to thank E. Brunet for explaining his results on the eigenstates of this model
and fruitful discussions. We are grateful to A. Borodin, P. Calabrese,
T. Gueudre, A. M. Povolotsky, J. Quastel, D. Remenik, T. Seppalainen, N. Zygouras for useful discussions. We thank I. Corwin for careful reading of the manuscript and useful remarks. This work is supported by a PSL grant ANR-10-IDEX-0001-02-PSL. 

\appendix

\section{Analytic continuation: Laplace transform from the moments}\label{appanalytic}

In this section, we illustrate the use of the Mellin-Barnes identity to compute the Laplace transform of a probability distribution
from its integer moments. In the most favorable cases the Laplace transform of the probability distribution $P(Z)$ of a positive random variable $Z$, such as a partition sum, can be calculated by a simple re-summation of the integer moments:
\bea \label{summ} 
 \overline{ e^{ - u Z} }  := \int_{Z>0} dZ P(Z) e^{- u Z}  = \sum_{n=0}^{+\infty} \frac{1}{n!} (-u)^n \overline{Z^n} 
\eea 

Clearly this formula cannot be used when some of the moments do not exist, e.g. when $P(Z)$ has an
algebraic tail. In that
case however one can use a more general formula in terms of a Mellin-Barnes transform. 

The basic identity is the following integral representation of the exponential function:
\begin{equation}
e^{-z} =\int_{-a + i \mathbb{R} }  \frac{ds}{2 i \pi} \Gamma(-s ) z^s = -\int_{-a + i \mathbb{R} }  \frac{ds}{2 i \sin(\pi s)} \frac{1}{\Gamma(1+s)} z^s 
\end{equation}
where $a>0$ and $z>0$. It allows us to express the Laplace transform of the probability distribution
$P(Z)$ as:
\begin{eqnarray}
&& \overline{ e^{ - u Z} }  
=  -  \int dZ P(Z)  \int_{-a + i \mathbb{R} }  \frac{ds}{2 i \sin(\pi s)} \frac{1}{\Gamma(1+s)} (uZ)^s \nonumber \\
&& =  -\int_{-a + i \mathbb{R} } \frac{ds}{2 i \sin(\pi s)} \frac{u^s}{\Gamma(1+s)}  \overline{ Z^s} \label{MB1} 
\end{eqnarray}
a more general formula, which is valid provided the integral converges. This is the case for
instance for the single site problem, i.e. $Z=w$ given by the inverse Gamma distribution, in which case
$\overline{w^s} = \Gamma(\gamma-s)/\Gamma(\gamma)$ for $Re(s) < \gamma$. In fact, in that
(trivial) case the formula (\ref{MB1}) is precisely the representation given in \cite{logsep2}, see e.g.
(\ref{sepp1}) setting $I=J=1$. 

\bigskip

In the case where $f(s) = \overline{Z^s}$ is analytic on the positive half-plane $Re(s) \geq 0$, and 
satisfies the conditions of Carlson theorem (i)  $ \exists $ $ C,\tau$, $|f(z)| < C e^{\tau z}$ 
(ii) $|f(i y)| < C e^{\pi y}$, the integral (\ref{MB1}) converges and we can close the contour on the positive
half plane. From the residues of the poles of the $1/\sin$ function one then recovers the 
formula (\ref{summ}) (equivalently, going from (\ref{summ}) to (\ref{MB1}) is nothing but the Mellin-Barnes
formula).

\section{Verifications of the formula for the norm}
\label{appgaudin}

Here we calculate the norm of the Brunet states in some simple cases, which provide verifications for the
general formula given in the text. 

\subsection{finite $L$}

For fixed $L$ one can directly compute the norm of a general 2 particles state with real momenta: $t_i = i \frac{k_i}{2} , k_i \in \mathbb{R} $. Using the formula for the weighted scalar product (\ref{wps}), one finds:
\begin{equation}
||\psi_{\mu}||^2 =  -\bar{c} L \frac{ 8 + k_1^2 + k_2^2}{(k_1-k_2)^2} + 2 L^2  \frac{ \left(\bar{c}^2+(k_1-k_2)^2\right)}{(k_1-k_2)^2} 
\end{equation}
in agreement with the formula (\ref{norme1}) using the modified Gaudin determinant. 

\subsection{in the limit $L \to +\infty$}
{\it Norm of a single $n$-string} 
In the limit $L \to \infty$, one can compute explicitly the norm of the state consisting of a single string (see section \ref{LargeL}), i.e. of particle content $m=n \in \mathbb{N}$. 
Inserting the string decomposition (\ref{string}) into the Brunet eigenfunctions (\ref{brunetansatz}), one sees that the single $n$-string eigenstate takes the simple form:
\bea \label{nstring}
\psi_{n-string}(x_1,\cdots,x_n) = n! z_1^{x_1} \cdots z_n^{x_n} \quad , \quad x_1 \leq \cdots \leq x_n
\eea
with $z_a = \frac{1+t_a}{1-t_a}$ and where the $t_a$ variables are organized as $t_a = i \frac{k}{2} +\frac{\bar{c}}{4} (m+1-2 a )$. 
For the infinite system one can recursively sum on the variables $y_i = x_i-x_{i-1}$ starting with $y_n$, carefully using 
the definition of the scalar product (\ref{wps}). Let us illustrate the calculation for $n=2,3$. One has:
\bea
||\psi_{2-string}||^2 = \sum_{x_1,x_2} \frac{1}{a_{x_1,x_2}}  |\psi_{2-string}(x_1,x_2)|^2 = 2 \sum_{x_1<x_2} 4 |z_1|^{2 x_1} |z_2|^{2 x_2} 
+ \frac{4}{h_2} \sum_{x_1} |z_1 z_2| ^{2 x_1} \nonumber \\
\simeq 8 L \sum_{y=1}^{+\infty} |z_2|^{2 y} + \frac{4 L}{h_2} 
\eea 
using $|z_1 z_2|=1$ from the Bethe equation. Using that $z_2=\frac{2 - \frac{\bar c}{2} + i k}{2 + \frac{\bar c}{2} - i k}$ 
one sees that $|z_2|<1$. Using that $h_2=4/(4-\bar c)$ and performing the sum one finds:
\begin{equation}
||\psi_{2-string}||^2 \simeq_{L \to \infty} \frac{L \left(4(4 + k^2) -\bar{c}^2\right)}{2 \bar{c}}
\end{equation}
in agreement with (\ref{norme2}).

A similar calculation for $n=3$ is performed using that
\bea
&& \sum_{x_1,x_2,x_3} a_{x_1,x_2,x_3}^{-1} |\psi(x_1,x_2,x_3)|^2 = 6 \sum_{x_1<x_2<x_3}   |\psi(x_1,x_2,x_3)|^2 \\
&& + \frac{3}{h_2} [ \sum_{x_1<x_3}
 |\psi(x_1,x_1,x_3)|^2 + \sum_{x_1<x_2}
 |\psi(x_1,x_2,x_2)|^2 ] + \frac{1}{h_3} \sum_{x_1}   |\psi(x_1,x_1,x_1)|^2
\eea
Inserting (\ref{nstring}), using that $|z_2|=1$, $|z_1|^2 = 1/|z_3|^2$ and $|z_3|^2 = \frac{(2-\bar c- i k)(2-\bar c+ i k)}{(2+\bar c- i k)(2+\bar c+ i k)}$
and performing the sums leads to the norm of the $3$-string as:
\begin{equation}
||\psi_{m=n=3}||^2 \sim_{L \to \infty} \frac{9 L \left( -16\bar{c}^2 + \bar{c}^4 + 3 (4+k^2)^2\right)}{8 \bar{c}^2}
\end{equation}
As one can see from this expression, it is hard to guess the general formula. Fortunately one can check that 
it agrees with the conjecture (\ref{norme2}).

\paragraph{}
{\it $n$ 1-strings:} 
In the case of $n$ particles with $n_s = n$, one easily obtains the norm in the large $L$ limit. 
In the calculation of $\sum_{x_1 , \cdots , x_n} \frac{1}{a_{x_1,..x_n}} \psi^*(x_1, \cdots , x_n) \psi(x_1, \cdots , x_n)$, one only encounters
plane waves with real momenta. It is then easy to see that, inserting the form (\ref{brunetansatz}) and expanding 
both wavefunctions in sum over permutations,
only the terms that come from the same permutation in $\psi^*$ and $\psi$ can give a power of $L^n$.
The computation of the other (non-diagonal) terms involve the use of the Bethe equations (\ref{BAE}) and give subdominant powers of $L$. Also, in that case, the factor $a_{x_1,..x_n}$
can be set to unity to leading order in the large $L$ limit. From there one easily obtains:
\bea\label{n1stringnorm}
|| \psi ||^2 =  n! L^n \prod_{i<j} \frac{ \bar{c}^2 + (k_i-k_j)^2}{(k_i-k_j)^2} + O(L^{n-1})  
\eea 
which is a consistency check of the first factor in the first formula (\ref{norme2}), and a check of the general norm formula (\ref{norme1}).

\section{Expansion of the eigenenergy around the LL limit}
\label{app:limits} 

Consider the expression for the eigenvalue (\ref{stringenergy}). The LL limit amounts to perform a small $\bar c$ expansion
at fixed $\tilde k=k/\bar c$. We can use the expansion of the Pochhammer symbol at 
at large $x$, $(x)_m = x^m f(x)$ with $f(x)=1 + \frac{m(m-1)}{2 x} + \frac{m(3 m^3 - 10 m^2 + 9 m -2)}{24 x^2} + O(1/x^3)$,
with $x=-\frac{m}{2} + \frac{\gamma}{2} + i \tilde k$ and $\gamma=1+\frac{4}{\bar c}$.
Then $\theta_{m,k}^2 = (\frac{2}{\bar c} |x|)^2 f(x) f(x^*)$, where $x^*$ is the complex conjugate. 
Since $\frac{2}{\bar c} |x| \to 1$ as $\bar c \to 0$ one can easily take the logarithm and expanding in $\bar c$,
up to $O(\bar c^4)$ one finds, up to terms of $O(\bar c^6,k^6,..)$:
\be 
- 8 \ln \theta_{m,k} = m k^2 + \frac{1}{12} (m-m^3) \bar c^2 
-\frac{\bar c^4 m \left(3 m^4-10
   m^2+7\right)}{1920}+\frac{1}{16} \bar c^2 k^2 m
   \left(m^2-1\right)-\frac{k^4 m}{8}
\ee
This expression is $O({\sf a}^2) + O({\sf a}^4)$ in the LL limit and when combined with the 
scaling of $t = \frac{t^{LL}}{8 {\sf a}^2} $ it gives the correct finite $LL$ limit displayed in the
text, together with the first correction in ${\sf a}$.

\section{Norm of strings from modified Gaudin formula in the limit $L \to \infty$}
\label{appgaudin2}

We start from the formula (\ref{norme1}) for the norm of an eigenstate 
given in the main text. As in the case of the Lieb-Liniger model, this formula is a-priori singular and the limit should be taken with care
for $L \to +\infty$ when string states appear. Here we follow the strategy of  \cite{cc-07}. In that limit
we split the $n$ particles into $n_s$ strings of multiplicity $m_j$: 
\begin{equation}
t_{j,a} = i \frac{ k_j}{2} + \frac{\bar{c}}{4} ( m_j + 1 - 2a ) +  \frac{\delta^{j,a}}{2} 
\end{equation}
where $j=1,...,n_s$ and $a=1,..., m_j$.
\paragraph{Limit of the prefactor in string notations:}
The prefactor is most conveniently written as 
\begin{equation}
\prod_{1 \leq \alpha < \beta \leq n }  \frac{   (2 t_{\alpha} - 2 t_{\beta} ) ^2 - \bar{c}^2  }   {   (2 t_{\alpha} - 2 t_{\beta} ) ^2  } =  \prod_{\alpha \neq \beta} \frac{  2 t_{\alpha} - 2 t_{\beta} -\bar{c}}{ 2 t_{\alpha} - 2 t_{\beta} }
\end{equation}
We now use the string notations and split the intra-string part from the inter-string part:
\begin{eqnarray}
 \prod_{\alpha \neq \beta} \frac{  2 t_{\alpha} - 2 t_{\beta}-\bar{c} }{  2t_{\alpha} - 2t_{\beta} } & = & \prod_{i \neq j} \prod_{a=1}^{m_i}\prod_{b=1}^{m_j} \frac{i( k_i - k_j) + \frac{\bar{c}}{2} (m_i - m_j -2(a-b+1))}{i( k_i - k_j) + \frac{\bar{c}}{2} (m_i - m_j -2(a-b))} \nonumber \\
 && \prod_{j=1}^{n_s} \prod_{a=1}^{m_j} \prod_{b \neq a} \frac{ \bar{c} (a-b+1) {\red - } \delta_{j}^{(a,b)}}{\bar{c} (a-b)}
\end{eqnarray}
where we denote $ \delta_{j}^{(a,b)} = \delta_{j,a}-\delta_{j,b}$ and keep these strings deviations only where needed for the
limit. After some work one finds that the leading term in the expansion in the strings deviations is given by:
 \bea
 \prod_{1\leq i < j  \leq n_s} \frac{4 (k_i-k_j)^2 + \bar{c}^2 (m_i + m_j)^2}{4 (k_i-k_j)^2 + \bar{c}^2 (m_i - m_j)^2}
\prod_{1 \leq j \leq n_s} m_j \left(\frac{1}{\bar{c}}\right)^{m_j-1} \prod_{a=1}^{m_j-1}  \delta_{j}^{(a,a+1)}
\eea

\paragraph{Limit of the modified Gaudin determinant:} 

Consider formula (\ref{modifiedgaudin}) in the main text. As in the Lieb-Liniger case, the determinant is singular and contains terms  of the form $K (t_{j,a} - t_{j,a+1}) =K_j^{(a,a+1)}   = \frac{1}{  \delta_{j}^{(a,a+1)}}  +O(1)$ that become exponentially large. It is easy to see that the leading term in the string deviation is obtained when one computes the determinant as if all string were decoupled: $\det G \sim \prod_{j=1}^{n_s} \det G_j$ where 
\begin{equation}
\det G_j =
\begin{vmatrix} 
L+(1-t_{j,1}^2)\sum_{b\neq 1} K_j^{(1,b)} & -(1-t_{j,1}^2)K_j^{(1,2)} & \cdots & -(1-t_{j,1}^2)K_j^{(1,m_j)}\\
-(1-t_{j,2}^2)K_j^{(1,2)} & L+(1-t_{j,2}^2)\sum_{b\neq 2} K_j^{(2,b)} & \cdots &-(1-t_{j,2}^2)K_j^{(2,m_j)} \\
. & . & \cdots & . \\
. & . & \cdots & . \\
-(1-t_{j,m_j}^2)K_j^{(1,m_j)}& -(1-t_{j,m_j}^2)K_j^{(2,m_j)} & \cdots & L+(1-t_{j,m_j}^2)\sum_{b\neq m_j} K_j^{(b,m_j)}
\end{vmatrix}
\end{equation}
This determinant can be handled in the same spirit as in \cite{cc-07}. One starts by adding the first column to the second one, then one adds to the second row the first one multiplied by $\frac{ 1-t_{j,2}^2}{1-t_{j,1}^2}$. The singular term $K_j^{(1,2)}$ now  only appears in the top-left entry and the entry $(2,2)$ now contains $L(1+ \frac{ 1-t_{j,2}^2}{1-t_{j,1}^2})$. One now iterates this procedure by adding the second column to the third one, and adding to the third row the second one multiplied by $\frac{ 1-t_{j,3}^2}{1-t_{j,2}^2}$, and the entry $(3,3)$ now contains $L\left(1 +\frac{ 1-t_{j,3}^2}{1-t_{j,2}^2}(1+ \frac{ 1-t_{j,2}^2}{1-t_{j,1}^2})\right) = L\left(1 +\frac{ 1-t_{j,3}^2}{1-t_{j,2}^2} + \frac{ 1-t_{j,3}^2}{1-t_{j,1}^2}\right)$. In the end all the singular terms $K_j^{(a,a+1)}$ are located on the first $m_j-1$ diagonal entries and the last term contains the leading power in $L$ which is $L(1-t_{j,m_j}^2) \sum_{b=1}^{m_j} \frac{1}{1-t_{j,b}^2}$. We thus obtain
\begin{equation}
\det G_j \sim L \left(\prod_{a=1}^{m_j-1} (1-t_{j,a}^2)K_j^{(a,a+1)} \right) (1-t_{j,m_j}^2) \sum_{b=1}^{m_j} \frac{1}{1-t_{j,b}^2}
\end{equation}
Note that we can do the exact same operation on the full modified Gaudin determinant to explicitly show that the different strings decouple. Taking all the strings into account, we thus arrive to:
\begin{equation}
\det{ G } \sim \prod_{j =1}^{n_s}  L  \left( \prod_{a=1}^{m_j-1}  \frac{1}{ \delta_{j}^{(a,a+1)}} \right) \prod_{a=1}^{m_j} (1-t_{j,a}^2) \sum_{b=1}^{m_j} \frac{1}{1-t_{j,b}^2}
\end{equation}
The divergent part precisely cancels the vanishing part of the prefactor and leads to the formula of the main text.

\section{Laplace transform versus moment generating function: some simple cases.}
\label{app:laplace} 

\paragraph{Calculations for the one-site problem $I=J=1$}

In the case of $Z=w$ distributed according to the inverse gamma distribution one can still close the contour in (\ref{MB1}). This coincides 
with the formula of \cite{logsep2} applied to one site. This leads to the result:
\bea
&& \overline{e^{- u Z}}  =  \sum_{n=0}^\infty \frac{(-u)^n}{n!}  \frac{\Gamma(\gamma-n)}{\Gamma(\gamma)} 
+ \sum_{n=0}^\infty \frac{(-1)^n}{n!}  u^{\gamma + n} \frac{\Gamma(- \gamma-n)}{\Gamma(\gamma)}  \\ 
&& = \frac{2}{\Gamma[\gamma] } u^{\frac{\gamma}{2}} K_{\gamma}(2 \sqrt{u})
\eea 
One can check that this is an exact formula. Notice that in the expansion, both sums converge separately but just give a part of the total Laplace transform:

\bea \label{nonanalyticexpansion}
&& \sum_{n=0}^\infty \frac{(-u)^n}{n!}  \frac{\Gamma(\gamma-n)}{\Gamma(\gamma)}  = u^{\gamma/2} \Gamma (1-\gamma) I_{-\gamma}\left(2 \sqrt{u}\right) \\
&& \sum_{n=0}^\infty \frac{(-1)^n}{n!}  u^{\gamma + n} \frac{\Gamma(- \gamma-n)}{\Gamma(\gamma)}  =  \frac{u^{\gamma/2} \Gamma (-\gamma) \Gamma (\gamma+1) I_\gamma\left(2
   \sqrt{u}\right)}{\Gamma (\gamma)} \nonumber
\eea 
where we used the usual notations for the Bessel functions. This is not apparent, but one can also notice that the sum of the (analytically-continued) moments possesses the symmetry $ \gamma \to 2- \gamma$, which is also the case for the Fredholm determinant computed in terms of hypergeometric functions computed in \ref{apphypergeom}. Note however that neither the Laplace transform, nor $P(w)$, possess this symmetry, another manifestation that the integer moments give only a part of the
total Laplace transform. The same property holds for the general case of arbitrary $t$, as discussed below. 

\paragraph{Calculation for t=2}

We now give a non-trivial check of the procedure for a length $2$ polymer. Consider the moments of $Z_{2}(0) = w_{0,0} ( w_{-\frac{1}{2},1}+w_{\frac{1}{2},1})w_{0,2} $ : they are given for $n< \gamma$ by

\begin{equation}
\overline{ Z_{2}(0)^n} = \sum_{k=0}^{n} C^{k}_{n} \frac{ \Gamma(\gamma -n)^2 \Gamma(\gamma -k) \Gamma(\gamma -(n-k)) }{\Gamma(\gamma) ^4}
\end{equation}
Because of the sum over $k$, it is not straightforward to analytically continue this formula in $n$. However, if we compute 
the moment generating function $g_{mom}(u)= \sum_{n=0} ^{\infty} (-1)^n \frac{u^n}{n!} \overline{ Z_{2}(0)^n}$, we obtain:

\bea
\!\!\!\!\!\!\!\!\!\!\!\!\!\!\!\!\!\!\!\! g_{mom}(u) = \sum_{ k_1\geq 0 , k_2 \geq 0} \frac{(-u)^{k_1+k_2}}{ \Gamma(1+ k_1 )   \Gamma( 1+k_2 ) } \frac{ \Gamma(\gamma -n)^2 \Gamma(\gamma -k) \Gamma(\gamma -(n-k)) }{\Gamma(\gamma) ^4}
\eea
On this function we can now perform the Mellin-Barnes trick to conjecture a formula for the Laplace transform $g(u)= \overline{ e^{ -u Z_{2}(0)} }$ :
\bea
&&  \!\!\!\!\!\!\!\!\!\!\!\!\!\!\!\!\!\!\!\! \!\!\!\!\!\!\! g(u)  = \frac{1}{4 \pi^2} \int_{-a + i \mathbb{R} } \int_{-a + i \mathbb{R} }  dk_1 dk_2 u^{k_1+k_2} \Gamma(-k_1 )   \Gamma( -k_2 ) \frac{ \Gamma(\gamma -n)^2 \Gamma(\gamma -k) \Gamma(\gamma -(n-k)) }{\Gamma(\gamma) ^4}  
\eea
where we used the reflection formula for the Gamma function. This formula is similar to the exact result obtained in \cite{logsep2}, and we
have numerically verified that the two results coincide. This provides a verification, for $t=2$ , of the general procedure detailed in the text
to conjecture the formula (\ref{Fredholmdet2}) for the Laplace transform for arbitrary $t$ using the Mellin-Barnes trick.

\section{ Generating Function as a Fredholm determinant}\label{Fredholmdet}

We start from the formula (\ref{Znsu}) for the partition sum at fixed number of strings. As in \cite{we} we use the following crucial identity:
\bea
\prod_{1\leq i < j  \leq n_s} \frac{4(k_i-k_j)^2 +  (m_i - m_j)^2}{4(k_i-k_j)^2 +  (m_i + m_j)^2}  = {\rm det}[ \frac{1}{2 i (k_i-k_j) + m_i + m_j} ]
\times \prod_{j=1}^{n_s} (2 m_j) 
\eea 
Hence we can rewrite  (\ref{Znsu}) as:
\bea
Z(n_s,u) & = &  \prod_{j=1}^{n_s} \sum_{m_j=1}^{+\infty} \int  \frac{dk_j}{ \pi}
 det[ \frac{1}{2 i (k_i-k_j) + m_i + m_j} ]  \nonumber \\
&& \times \prod_{j=1}^{n_s} (-u)^{m_j}   \prod_{j=1}^{n_s} 
\left( \frac{  \Gamma(-\frac{m_j}{2} + \frac{ \gamma}{2} - i k_j ) }{  \Gamma(\frac{m_j}{2} + \frac{ \gamma}{2} - i k_j )  } \right)^{I} \left( \frac{   \Gamma(-\frac{m_j}{2} + \frac{ \gamma}{2} +i k_j )}{ \Gamma(\frac{m_j}{2} + \frac{ \gamma}{2} + i k_j ) } \right)^{J}
\eea 
The determinant can be written as a sum over permutations $\sigma$, and we also introduce the representation $\frac{1}{x} =  \int_{ R_+} dv e^{-v x}$, which leads to
\bea
 Z(n_s,u) & = & \sum_{\sigma \in S_n} (-1)^\sigma \prod_{j=1}^{n_s} \sum_{m_j=1}^{+\infty} \int  \frac{dk_j}{ \pi} \int_{v_j>0} 
e^{- v_j ( 2 i (k_j -k_{\sigma(j)} + m_j +m_{\sigma(j)}) } (-u)^{m_j}  \nonumber \\
&& \times   \prod_{j=1}^{n_s} 
\left( \frac{  \Gamma(-\frac{m_j}{2} + \frac{ \gamma}{2} - i k_j ) }{  \Gamma(\frac{m_j}{2} + \frac{ \gamma}{2} - i k_j )  } \right)^{I} \left( \frac{   \Gamma(-\frac{m_j}{2} + \frac{ \gamma}{2} +i k_j )}{ \Gamma(\frac{m_j}{2} + \frac{ \gamma}{2} + i k_j ) } \right)^{J}\nonumber
\eea 
We then perform the change $\sum_j v_j k_{\sigma(j)} = \sum_j v_{\sigma^{-1}(j)} k_j$ ( and the same for $\sum_j v_j m_{\sigma(j)} $) and relabel as $\sigma \to \sigma^{-1}$, this leads to:
\bea
Z(n_s,u) &=&  \sum_{\sigma \in S_n} (-1)^\sigma \prod_{j=1}^{n_s} \sum_{m_j=1}^{+\infty} \int  \frac{dk_j}{ \pi} \int_{v_j>0} 
e^{- 2 i k_j (v_j -v_{\sigma(j)}) -  m_j (v_j + v_{\sigma(j)}) } (-u)^{m_j}  \nonumber \\
&& \left( \frac{  \Gamma(-\frac{m_j}{2} + \frac{ \gamma}{2} - i k_j ) }{  \Gamma(\frac{m_j}{2} + \frac{ \gamma}{2} - i k_j )  } \right)^{I} \left( \frac{   \Gamma(-\frac{m_j}{2} + \frac{ \gamma}{2} +i k_j )}{ \Gamma(\frac{m_j}{2} + \frac{ \gamma}{2} + i k_j ) } \right)^{J}  \nonumber
\eea 
which has the structure of a determinant:
\begin{equation}
Z(n_s,u) = \prod_{j=1}^{n_s}  \int_{v_j>0}  {\rm det}[ K^{mom}_{I,J}(v_i,v_j) ]_{n_s \times n_s} 
\end{equation} 
with the kernel $K^{mom}_{I,J}$ given in (\ref{firstfredholm}). Summation over $n_s$ leads to the
Fredholm determinant expression given in the text. 

%

\section{Moments-kernel in term of hypergeometric functions}\label{apphypergeom}

We show that the moments-kernel $K_{mom}$ can be exactly expressed in terms of hypergeometric functions by separating the
summation over $m$ even and $m$ odd. We restrict to $t$ even and $x=0$ and define:
\begin{eqnarray}
\fl && G_n(k,z) =  \sum_{m=1}^{\infty} (-z)^m \left( \frac{ \Gamma( -\frac{m}{2} + \frac{\gamma}{2} - ik ) \Gamma( -\frac{m}{2} + \frac{\gamma}{2} + ik ) }{ \Gamma( \frac{m}{2} + \frac{\gamma}{2} - ik ) \Gamma( \frac{m}{2} + \frac{\gamma}{2} + ik ) } \right) ^{n} = -1 + A_n(k,z^2) -z B_n(k,z^2) \nonumber
\end{eqnarray}
with
\begin{equation}
A_n(k,z ) = \sum_{ m = 0}^{\infty} z^m \left( \frac{ \Gamma( -m + \frac{\gamma}{2} - ik ) \Gamma( -m + \frac{\gamma}{2} + ik ) }{ \Gamma( m + \frac{\gamma}{2} - ik ) \Gamma( m + \frac{\gamma}{2} + ik ) } \right) ^{n}
\end{equation}
and
\begin{equation}
B_n(k,z ) = \sum_{ m = 0}^{\infty} z^m \left( \frac{ \Gamma( -m - \frac{1}{2}  + \frac{\gamma}{2} - ik ) \Gamma( -m -\frac{1}{2} + \frac{\gamma}{2} + ik ) }{ \Gamma( m + \frac{1}{2} + \frac{\gamma}{2} - ik ) \Gamma( m + \frac{1}{2} + \frac{\gamma}{2} + ik ) } \right) ^{n}
\end{equation}
Using the Euler reflection formula three times, we obtain:
\begin{eqnarray}
 && \!\!\!\!\!\!\!\!\!\!\!\!\!\!\!\!\!\!\!\!\!\!\!\!\!\!\! \Gamma( -m + \frac{\gamma}{2} - ik ) \Gamma( -m + \frac{\gamma}{2} + ik ) =  \frac{ \Gamma(\frac{\gamma}{2} - ik ) \Gamma( 1 - \frac{\gamma}{2} + ik ) \Gamma( \frac{\gamma}{2} + ik ) \Gamma( 1 - \frac{\gamma}{2} - ik )}{ \Gamma(1 +m - \frac{\gamma}{2} + ik ) \Gamma( 1+m - \frac{\gamma}{2} - ik ) } \nonumber 
\end{eqnarray}
This allows to express
\begin{eqnarray}
A_n(k,z ) & = & _1F_{4n} \left(\{1\} , \{ ( 1 - \frac{\gamma}{2} + ik ) ,  ( 1 - \frac{\gamma}{2} - ik ), (\frac{\gamma}{2} - ik ) ,  ( \frac{\gamma}{2} + ik ) \}_n  ; z \right)  
\end{eqnarray}
where we denote:
\bea
&&  \{ ( 1 - \frac{\gamma}{2} + ik ) ,  ( 1 - \frac{\gamma}{2} - ik ), (\frac{\gamma}{2} - ik ) ,  ( \frac{\gamma}{2} + ik ) \}_n= \\
&& \bigoplus_{i=1}^n \{ ( 1 - \frac{\gamma}{2} + ik ) ,  ( 1 - \frac{\gamma}{2} - ik ), (\frac{\gamma}{2} - ik ) ,  ( \frac{\gamma}{2} + ik ) \} \nonumber
\eea
The same type of calculation leads to
\begin{eqnarray}
&& B_n(k,z )  =    \left( \frac{4}{ (\gamma-1)^2 + 4 k^2  }  \right)^n  \\
&&   _{1}F_{4n} \left(\{1\} ,\{ ( \frac{3}{2} - \frac{\gamma}{2} + ik ) ,  ( \frac{3}{2} - \frac{\gamma}{2} - ik ), ( \frac{1}{2} + \frac{\gamma}{2} - ik ) ,  (  \frac{1}{2} + \frac{\gamma}{2} + ik ) \}_n  ; z \right) \nonumber
\end{eqnarray}
And this allows to express $K_{mom}$ in (\ref{firstfredholm}) as:
\begin{equation}
K^{mom}(v_1,v_2) = \int_{ \mathbb{R}} \frac{dk}{\pi} e^{ -  2 i k(v_1-v_2)} \left( -1 + A_{ \frac{t}{2}+1} ( k , u^2 e^{ - 2 (v_1 + v_2 ) } ) - u e^{ -  (v_1 + v_2 ) }    B_{ \frac{t}{2}+1} ( k , u^2 e^{  - 2 (v_1 + v_2 ) } )  \right)
\end{equation}
The interesting feature is that on this result, the symmetry $\gamma \to 2-\gamma$ holds. Since we know that the Laplace transform
cannot have this symmetry, this shows once again that it cannot be equal to the moment generating function.

\section{Some verifications of the various kernels}\label{appcheck}

For $t$ even and $x=0$ (centered arrival point), the kernel (\ref{firstfredholm}) takes the form
\begin{eqnarray}
&&  K^{mom}_t(v_1,v_2) =  \\
  && \!\!\!\!\!\!\!\!\!\!\!\!\!\!\!\!\!  \sum_{m=1}^{\infty} \int_{-\infty}^{+\infty}   \frac{dk}{ \pi}  (-1)^m u^m  e^{ -  2 i k(v_1-v_2) -  m (v_1+v_2) }  \left( \frac{ \Gamma( -\frac{m}{2} + \frac{\gamma}{2} - ik ) \Gamma( -\frac{m}{2} + \frac{\gamma}{2} + ik ) }{ \Gamma( \frac{m}{2} + \frac{\gamma}{2} - ik ) \Gamma( \frac{m}{2} + \frac{\gamma}{2} + ik ) } \right)^{\frac{t}{2}+1} \nonumber
\end{eqnarray}
The integration over $k$ can be performed by noting that 
there are two series of poles $i k =\pm(  - p + \frac{m-\gamma}{2})$, $p \in \mathbb{N}$, in the gamma functions (the use of the residues formula here is legitimate, since, as in the main text, one can easily rewrite the quotient of Gamma functions as a rational fraction).

Consider $t=0$. Let us consider for now only the terms $m < \gamma$, our goal will be to recover the moments
$n<\gamma$ from the Fredholm determinant. The integral over $k$ can be performed by closing 
the contour on the side $i k>0$ or $i k<0$ depending on the sign of $v_1-v_2$ leading to:
\bea
K^{mom}_{t=0}(v_1,v_2) =   2  \sum_{m=1}^{\infty} \sum_{p=0}^{m-1} \frac{(-1)^p}{p!} \frac{\Gamma(\gamma+p-m)}{\Gamma(m-p) \Gamma(\gamma+p)} 
(-u)^{m} e^{- (2 p + \gamma - m) |v_1-v_2| - m (v_1+v_2)} \nn
\eea 
since for $m<\gamma$ one picks either the first series of poles $i k>0$ or the second.

Here at $t=0$, we want to check that:
\bea
 {\rm Det} \left( I  + K^{mom}_{t=0} \right)|_{(-u)^n} = n! \Gamma(\gamma-n)/\Gamma(\gamma) 
\eea 
We can use the expansion:
\bea
&& {\rm Det} ( I  + K) = e^{  \Tr \ln(I + K) } = 1  + \Tr K + \frac{1}{2} ( (\Tr K)^2 - \Tr K^2)  \\
&&  + \frac{1}{6} ((\Tr K)^3 - 3 \Tr K  \Tr K^2 + 2 \Tr K^3) + .. \nonumber
\eea
we now denote $K=\sum_m K_m$ and check up to order 3 or 4 ..

The same reasoning can be applied to the different kernels obtained from this one in the text.
One can check that (\ref{Fredholmdet2}) and (\ref{finalkernel}) indeed give the moments of the distribution (checked at $t=0$ and $t=2$).
One can also check that the first non-analytic terms in the Laplace transform of the probability distribution at $t=0$ are reproduced. For that one
starts from 
(\ref{finalkernel}) and explicitly calculate the integral over $w$ using residues
\begin{eqnarray}
&&  K^{BA}_{1,1}(z,z') =\frac{1}{2\pi i} \sum_{n_1=1}^{\infty} \frac{(-u)^{n_1}}{z+n_1-z'}  \frac{ \Gamma( \alpha- z-n_1) \Gamma( z - a)}{\Gamma( z+n_1 - a) \Gamma( \alpha - z)} \\
 &+&\frac{1}{2\pi i} \sum_{n_2=0}^{\infty} \frac{\pi}{\sin(\pi(z-\alpha-n_2))} \frac{(-1)^{n_2}}{n_2!} \frac{u^{\alpha+n_2-z}}{\alpha+n_2-z'}  \frac{ \Gamma( z - a)}{\Gamma( \alpha+n_2
 - a) \Gamma( \alpha - z)} \nonumber
 \end{eqnarray}
Using this expansion allows to recover the first terms in \ref{nonanalyticexpansion} and in particular the non analytic terms $\frac{(-1)^n}{n!} u^{\gamma+n} \frac{\Gamma(-\gamma-n)}{\Gamma(\gamma)}$ (we checked it for $n=0,1$). The various traces can be computed using the residues theorem. Integer powers of $u$ come from the first part of the expansion and from the poles of the sine function in the second part, whereas non-integer powers of $u$ come from the poles of the Gamma function in the second part. The fact that we can extract the correct integer moments from the kernels is a consistency check of the procedure. On the other hand, being able to retrieve the non analyticity is another sign that the Mellin-Barnes trick indeed provides the correct analytic continuation.

\section{Probability distribution at any time}\label{app:proba}

Starting from the expression for the generating function $g_{I,J}(u) = \overline{ e^{ - u Z(I,J) } }$ and writing formally $Z(I,J)$ as the product of a variable $Z_0$ with an exponential distribution: $P_0(Z_0) = e^{-Z_0}$ (i.e. $\log Z_0$ has a unit Gumbel distribution), and a new positive random variable $\tilde{Z}(I,J)$ distributed according to $\tilde{P}_{I,J}$, one has
\begin{equation}
g_{I,J}(u) = \overline{e^{- u Z_0 \tilde Z(I,J) } }=\overline{\int dZ_0 e^{ - u Z_0 \tilde Z(I,J) } e^{-Z_0} }=\overline{ \frac{1}{1+u \tilde{Z}(I,J)} } = \int d\tilde{Z} \frac{1}{1+ u \tilde{Z}} \tilde{P}_{I,J}(\tilde{Z}) 
\end{equation}
Assuming an analytic continuation, we write
\begin{equation}
g_{I,J}(\frac{1}{-v-i\epsilon}) = \int d\tilde{Z} \frac{-v }{ \tilde{Z}-v - i \epsilon} \tilde{P}_{I,J}(\tilde{Z}) 
\end{equation}
And the limit $\epsilon \to 0^+$ 
allows to extract the probability distribution $\tilde{P}_{I,J}$ as
\begin{equation}
\tilde{P}_{I,J}(v) = \frac{1}{2 i \pi v } \lim_{\epsilon \to 0^+} \left( g_{I,J}(\frac{1}{-v+i\epsilon})-g_{I,J}(\frac{1}{-v-i\epsilon}) \right) 
\end{equation}
Using (\ref{Fredholmdet2}), we write $g_{I,J}(\frac{1}{-v\pm i\epsilon}) = {\rm Det} (I+\check{K}_{I,J}^\pm)$ with
\begin{eqnarray}
\check{K}_{I,J}^\pm(v_1,v_2) = && \int_{-\infty}^{+\infty}   \frac{dk}{ \pi}  \frac{-1}{2i} \int_C \frac{ds}{ \sin( \pi s ) }   \left( \frac{1}{-v \pm i \epsilon} \right)^s  e^{ -  2 i k(v_1-v_2) -  s (v_1+v_2) } \\
 &&  \left( \frac{  \Gamma(-\frac{s}{2} + \frac{ \gamma}{2} - i k ) }{  \Gamma(\frac{s}{2} + \frac{ \gamma}{2} - i k )  } \right)^{I} \left( \frac{   \Gamma(-\frac{s}{2} + \frac{ \gamma}{2} +i k )}{ \Gamma(\frac{s}{2} + \frac{ \gamma}{2} + i k ) } \right)^{J} \nonumber 
\end{eqnarray}
Using the principal determination of the logarithm, and since $v$ has to be positive, we have
\begin{equation}
\lim_{\epsilon \to 0^+} \left( \frac{1}{-v \pm i \epsilon} \right)^s = \exp( - s \log(v)  \mp i \pi s)
\end{equation}
Finally, writing $e^{\mp i \pi s} = \cos(\pi s) \mp i \sin(\pi s)$ leads to the formula of the main text.

\section{Saddle point position}\label{appnumerics}

The numerical resolution of the saddle-point equation (\ref{saddlepoint}), i.e.:
\be
\frac{\frac{1}{2} + \varphi}{\frac{1}{2} - \varphi} = \frac{\psi'(\frac{\gamma}{2} - k_{\varphi})}{\psi'(\frac{\gamma}{2} + k_{\varphi})}
\ee
is complicated by the divergence near $\varphi=\frac{1}{2}$. In fact there is a solution such that the 
argument of the $\psi'$ function remains positive. Since $\lim_{x \to 0^+} \psi'(x) = +\infty$ 
it is easy to see that $\lim_{\varphi \to \frac{1}{2}} k_\varphi = -\frac{\gamma}{2} $. 
Explicitly, the leading behavior of $k_{\varphi}$ is
\begin{equation}
k_{\varphi}\simeq_{\varphi \to \frac{1}{2}} -\frac{\gamma}{2} + \left( \frac{ \frac{1}{2} - \varphi}{ \psi'(\gamma)} \right)^{\frac{1}{2}} +... 
\end{equation}
This divergence makes the numerical solution fail around $\varphi = \frac{1}{2}$: $k_\varphi$ crosses the singularity at $-\frac{\gamma}{2}$. On the other hand, the non-analyticity makes a perturbative calculation inefficient close to this point. The most accurate determination appears to be a fit between the numerical result and the known non analyticity, which is what was used for Fig. \ref{figfullconvergence} and \ref{figfullspatialdependance} in the text. Fig \ref{figsaddlepoint} summarize the situation.
\begin{figure}[H]
   \centering
   \includegraphics[scale=0.4]{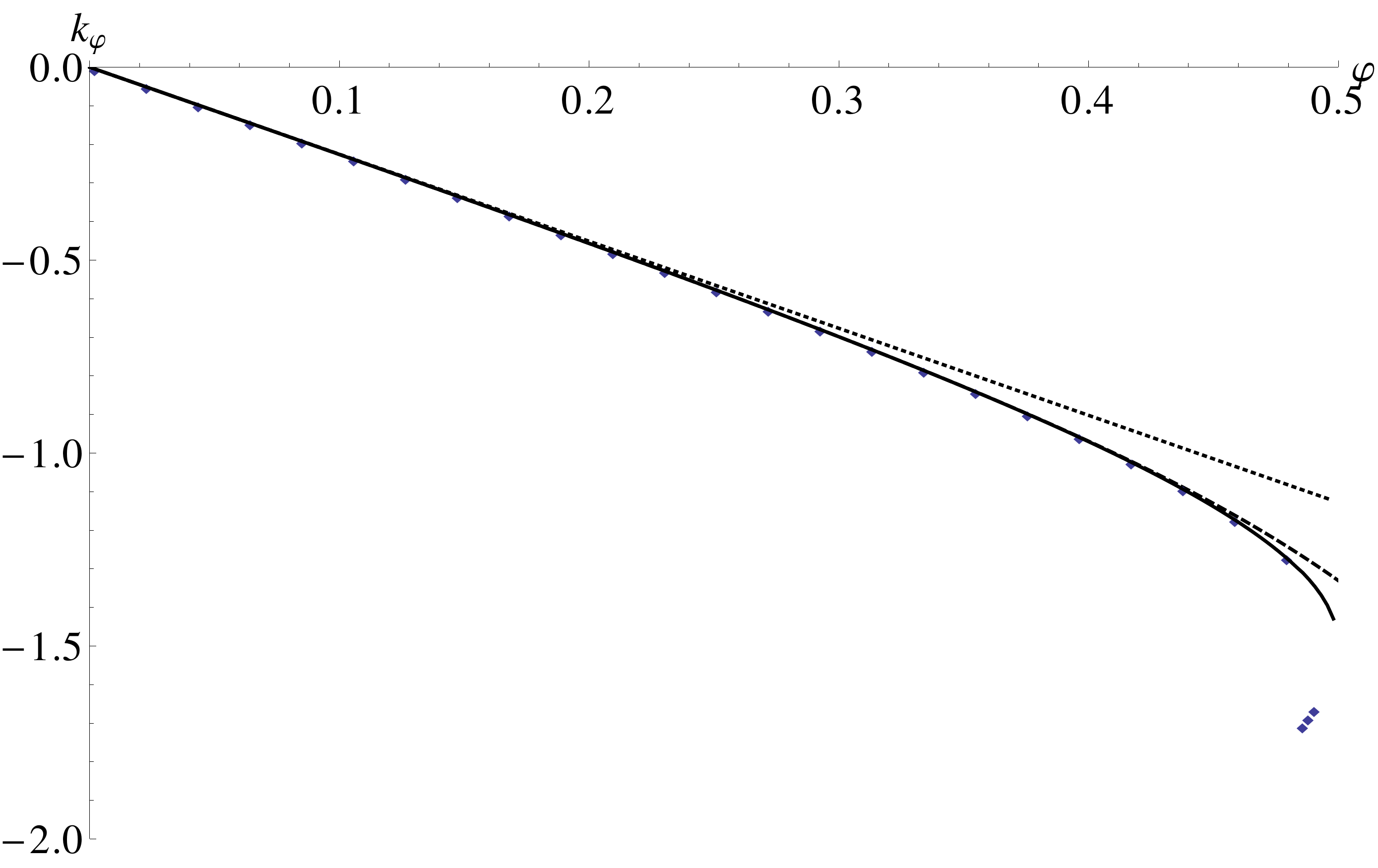}
   \caption{Saddle-point position $k_{\varphi}$ as a function of $\varphi$ for $\gamma=3$. The dotted-line is the approximation to lowest order
   in $\varphi$, i.e.  $k_{\varphi} \sim \varphi$ (STS). The losanges are the numerical solution. The dashed line is a high order perturbative approximation and the solid line is the final result that uses the non-analytic behaviour near $\varphi=\frac{1}{2}$. The additional
   points below arise from numerical artefacts.}
   \label{figsaddlepoint}
\end{figure}
\paragraph{}

\section{The semi-directed random polymer}\label{appsemi}

The semi-directed random polymer was introduced by O'Connell and Yor in \cite{OConnellYor,semidiscret1}. In  \cite{semidiscret2}  
it was argued that it constitutes an universal scaling limit for polymer restricted to stay close to the boundary (with proper rescaling of the temperature or in our case, of the parameter of the inverse-gamma distribution). In the simplest case (no drift, temperature and total polymer length $t$ set to unity) it is defined as the partition sum

\bea
Z_N^{s.d.}= \int_{0 < s_1< \cdots <s_{N-1} \leq 1} e^{B_1(s_1) + (B_2(s_2)-B_2(s_1))+\cdots + (B_N(1)-B_N(s_{N-1}))}
\eea
where $B_j(s)$ are $N$ independent standard Brownian motions. 
In \cite{logsep2}, it was shown that this model could be obtained as the following scaling limit of the log-Gamma polymer: $Z_N^{s.d.} \sim_{in law} \lim_{n \to \infty} e^{ n \log(n) - \frac{1}{2}} Z(n , N)|_{\gamma=n}$. Here we show how this scaling limit naturally appears and we obtain a Fredholm-Determinant formula for the Laplace transform of the semi-directed polymer partition sum.
Starting from (\ref{finalkernel}) 
we need to analyze the large $n$ limit of ${\rm Det} (I+K_{n,N}^{BA})$ where
\begin{equation}
K_{n,N}^{BA}(z , z') = \int_{ a+i \mathbb{R} } d s 
\frac{1}{4 \pi (  s+z-z')} \frac{1}{ \sin( \pi s) }   u^{  s }  \left( \frac{ \Gamma(  z  ) }{  \Gamma( z+s ) }\right) ^{ N} \left( \frac{\Gamma(n - z-s ) }{ \Gamma(n-z) } \right) ^{ n  }    
\end{equation}
and $z,z' \in \tilde a + i \mathbb{R}$. We have defined $s=w-z$ and renamed $z-a \to z$. Here the factor $\left( \frac{\Gamma(n - z-s ) }{ \Gamma(n-z) } \right) ^{ n  }$ takes a simple form in the large $n$ limit:
\bea
\left( \frac{\Gamma(n - z-s ) }{ \Gamma(n-z) } \right) ^{ n  } = \exp\left( n\left( -s \psi(n) +\frac{1}{2}\psi'(n) (2 s z + s^2) + O(\frac{1}{n^2}) \right) \right)
\eea
where use that $\psi^{(k)}(n) = O( \frac{1}{n^k})$ for $n\to \infty$. Using $\psi(n) =_{n \to \infty} \log(n) - \frac{1}{2 n } + O(\frac{1}{n^2})$ and $\psi'(n) =_{n \to \infty} \frac{1}{ n } + O(\frac{1}{n^2})$, we thus arrive at:
\bea
\left( \frac{\Gamma(n - z-s ) }{ \Gamma(n-z) } \right) ^{ n  } \sim_{n\to\infty} \exp\left( -s (n \log n -\frac{1}{2}) +s z+\frac{1}{2} s^2\right)
\eea
The first term indeed imposes to rescale the partition sum as $\hat{Z}(n,N) = e^{  n \log(n) - \frac{1}{2}} Z(n,N)$ so that the laplace transform of $\hat{Z}(n,N)$, $\hat{g}_{n,N} = \overline{ \exp{-u\hat{Z}(n,N)}} $ has a well-defined $n \to \infty$ limit given by a Fredholm determinant, with:
\bea
&& \lim_{n \to +\infty} \hat{g}_{n,N}  =  {\rm Det}( I+\hat{K}_N) \\
&& \hat{K}_N(z,z') = \int_{ a + i \mathbb{R} } d s 
\frac{1}{4 \pi ( s+z-z')} \frac{1}{ \sin( \pi s) }   u^{  s }  \left( \frac{ \Gamma(  z  ) }{  \Gamma( z+s ) }\right) ^{ N} e^{ s z + \frac{1}{2} s^2} 
\eea
and $z,z' \in \tilde a + i \mathbb{R}$. We recall $0<a<1$ and $0<\tilde a$ (in the limit). 
This result is identical to Theorem 3 of \cite{logboro} for the case of zero drift and $t=1$
(see also Theorem 1.5 in \cite{corwinsmallreview}) 
apart from the (now usual) difference of contours. There $z,z'$ belong 
to a small circle around 0, while the $s$ contour is the same. A similar (large-contour) formula can be found in Theorem 1.17. of \cite{BCF}. There (for our case), the contour of integration on $z$  is a wedge $C_{\alpha , \phi} = \{ \alpha +e^{i (\pi+\phi)} \mathbb{R}_+ \} \cup \{ \alpha +e^{i (\pi-\phi)} \mathbb{R}_+ \}   $ where $\alpha >0$ and $0<\phi<\pi/4$, and the contour of integration on $s$, $D_z$, is $z$-dependent and given by straight-lines joining $R(z) - i \infty$ to $R(z)-id$ to $\frac{1}{2}-id$ to $\frac{1}{2}+id$ to $R(z)+id$ to $R(z) + i \infty$, where $R(z) = -Re(z)+\alpha+1$ and $d>0$ is small enough so that to ensure that $z +D_z$ do not intersect $C_{\alpha , \phi}$. These contours are more involved but are similarly located as ours with respect to the poles of the integrand.


\section*{References}




\begin{thebibliography}{32}


\bibitem{logsep1}
T. Sepp\"{a}l\"{a}inen. Scaling for a one-dimensional directed polymer with boundary. Ann. Probab., {\bf 40} :19-73, (2012).


\bibitem{logsep2}
I. Corwin, N. OÃ¢ÂÂConnell, T. Sepp\"{a}l\"{a}inen, and N. Zygouras. Tropical combinatorics and Whittaker functions.
arXiv:1110.3489 (2011).


\bibitem{logboro}
Borodin, A., Corwin, I., Remenik, D.: Log-gamma polymer free energy Ã¯Â¬Âuctuations via a Fredholm
determinant identity. Comm. Math. Phys. {\bf 324} , 1 , 215-232 (2013).

\bibitem{KPZ}
M. Kardar, G. Parisi and Y.C. Zhang, Phys. Rev. Lett. {\bf 56}, 889 (1986).

 \bibitem{TW1994}
C.A. Tracy and H. Widom, Comm. Math. Phys. {\bf 159}, 151 (1994).

\bibitem{kardareplica}
M. Kardar, Nucl. Phys. B {\bf 290}, 582 (1987).

\bibitem{ll} E. H. Lieb and W. Liniger, Phys. Rev. {\bf 130}, 1605 (1963).

\bibitem{we}
P. Calabrese, P. Le Doussal and A. Rosso, EPL {\bf 90}, 20002 (2010). 

\bibitem{dotsenko}
V. Dotsenko, EPL {\bf 90}, 20003 (2010); J. Stat. Mech. P07010 (2010);  
V. Dotsenko and B. Klumov, J. Stat. Mech. (2010) P03022.

\bibitem{we-flat}
P. Calabrese and P. Le Doussal, Phys. Rev. Lett. {\bf 106}, 250603 (2011); 
P. Le Doussal and P. Calabrese,  J. Stat. Mech.  P06001 (2012).

\bibitem{SasamotoStationary}
T. Imamura, T. Sasamoto, arXiv:1111.4634, 
Phys. Rev. Lett. {\bf 108}, 190603 (2012); and arXiv:1105.4659,
J. Phys. A  {\bf 44},  385001 (2011). 

\bibitem{dg-12}
T. Gueudr\'e and P. Le Doussal,  EPL {\bf 100}, 26006 (2012).

\bibitem{d-13}
V. Dotsenko, J. Phys. A {\bf 46}, 355001 (2013);  J. Stat. Mech. (2013) P06017; J. Stat. Mech. (2013) P02012.

\bibitem{psn-11}
T Imamura, T. Sasamoto, and H. Spohn,  J. Phys. A {\bf 46}, 355002 (2013);
S. Prolhac and H. Spohn, J. Stat. Mech. (2011) P01031, J. Stat. Mech. (2011) P03020.

\bibitem{ld-14}
P. Le Doussal, J. Stat. Mech. (2014) P04018. 

\bibitem{cl-14}
P. Calabrese and P. Le Doussal, arXiv:1402.1278. 

\bibitem{sineG}
P. Calabrese, M. Kormos and P. Le Doussal, arXiv:1405.2582.

\bibitem{spohnKPZEdge}
T. Sasamoto and H. Spohn, Phys. Rev. Lett. {\bf 104}, 230602 (2010),
Nucl. Phys. B {\bf 834}, 523 (2010), J. Stat. Phys. {\bf 140}, 209 (2010).

\bibitem{corwinDP}
G. Amir, I. Corwin, J. Quastel, Comm. Pure Appl. Math {\bf 64}, 466 (2011);
I. Corwin, arXiv:1106.1596. 

\bibitem{reviewCorwin}
I. Corwin, Random Matrices Theory Appl., 1, 2012, arXiv:1106.1596 

\bibitem{BorodinMacdo}
A. Borodin and I. Corwin Prob. Theor. Rel. Fields 158 (2014), no. 1-2, 225--400, arXiv:1111.4408

\bibitem{BorodinQboson}
A. Borodin, I. Corwin, L. Petrov, T. Sasamoto, arXiv:1308.3475.

\bibitem{BCS} 
A. Borodin, I. Corwin, T. Sasamoto. From duality to determinants for q-TASEP and ASEP. Ann. Probab., to appear. arXiv:1207.5035.

\bibitem{BCF}
A. Borodin, I. Corwin, P. L. Ferrari. Free energy fluctuations for directed polymers in random media in 1+ 1 dimension. Comm. Pure Appl. Math., to appear. arXiv:1204.1024.

\bibitem{quastel-prep}
 J. Quastel, J. Ortmann, and D. Remenik in preparation.

\bibitem{Johansson2000} K. Johansson, Comm. Math. Phys. {\bf 209}, 437 (2000).

\bibitem{semidiscret1}
N. O'Connell. Directed polymers and the quantum Toda lattice. Ann. Probab., {\bf 40}:437Ã¢ÂÂ458, (2012).

\bibitem{semidiscret2} A. Auffinger, J. Baik, I. Corwin. Universality for directed polymers in thin rectangles. arXiv:1204.4445 (2012).

\bibitem{corwinsmallreview} I. Corwin, Proceedings of the ICM, arXiv:1403.6877.

\bibitem{Brunet1}
E. Brunet, Private communication and to be published. 

\bibitem{m-65} J. B. McGuire, J. Math. Phys. {\bf 5}, 622 (1964).

\bibitem{cc-07} 
P. Calabrese and J.-S. Caux, Phys. Rev. Lett. {\bf 98}, 150403 (2007); J. Stat. Mech. (2007) P08032.

\bibitem{bb-00}
E. Brunet and B. Derrida, Phys. Rev. E {\bf 61}, 6789 (2000);  Physica A {\bf 279}, 395 (2000). 

\bibitem{brunet}
Brunet E., PhD Thesis.
   
\bibitem{gk}
M. Gaudin, La fonction d'onde de Bethe, Masson, Paris, 1983; 

\bibitem{OConnellYor}
O'Connell, N. and Yor, M. (2001). 
Stochastic Process. Appl. 96 285Ð304.

\bibitem{povolo} A.M. Povolotsky. On integrability of zero-range chipping models with factorized steady state. J. Phys. A,
{\bf 46}:465205, 2013.

\bibitem{Borodinprivate} 
We thank A. Borodin and I. Corwin for a discussion on this point and on their upcoming work
\cite{BCFV}.

\bibitem{BCFV}
A. Borodin, I. Corwin, P. L. Ferrari. B. Veto. Height fluctuations for the stationary KPZ equation, in preparation.

\bibitem{foot4} Using that $| \Gamma(x+i y) | \simeq \sqrt{2 \pi} |y|^{x-\frac{1}{2}} e^{- \frac{\pi}{2} |y|}$ for
$x,y \in \mathbb{R}$ and $|y| \to +\infty$.

\end{thebibliography}
\end{document}